\documentclass[superscriptaddress,nofootinbib,aps,twocolumn]{revtex4-2}

\usepackage[utf8]{inputenc}
\usepackage{typearea,amsmath,amsthm,color,typearea,multirow,graphicx}
\usepackage[braket, qm]{qcircuit}
\renewcommand{\vec}[1]{{\boldsymbol{\bf #1} }}
\newcommand{\dataspace}[0]{\mathcal{X}}
\newcommand{\expect}[2]{\mathbf{E}_{#1}\left[#2\right]}

\newcommand{\dhs}[0]{d^{\rm HS}(\rhoq{\vec{x}}, \rhoq{\vec{x}^{\prime}})}
\newcommand{\dhsk}[0]{d^{\rm HS}(\rhoqk{\vec{x}}, \rhoqk{\vec{x}^{\prime}})}
\newcommand{\pqj}[0]{P_j^{(\vec{q})}}
\newcommand{\pqjk}[0]{P_j^{(\vec{q}_k)}}
\newcommand{\pqjprime}[0]{P_{j^{\prime}}^{(\vec{q})}}
\newcommand{\pqji}[0]{(\pqj \otimes \iq)}
\newcommand{\pqjki}[0]{(\pqjk \otimes \iqk)}
\newcommand{\iq}[0]{I^{(\neq \vec{q})}}
\newcommand{\iqk}[0]{I^{(\neq \vec{q}_k)}}
\newcommand{\ns}[0]{N_{\rm shot}}
\newcommand{\rhoq}[1]{\rho^{(\vec{q})}(#1)}
\newcommand{\rhoqk}[1]{\rho^{(\vec{q}_k)}(#1)}
\newcommand{\tildepq}[0]{\tilde{P}_j^{(\vec{q})}}
\newcommand{\Nq}[0]{N_{\vec{q}}}

\newcounter{todo}
\newtheorem{th.}{Theorem}
\newtheorem{co.}{Corollary}
\newtheorem{definition}{Definition}

\renewcommand{\r}[0]{\mathbf{R}}
\typearea{17}

\begin{document}
\title{Deterministic and random features for large-scale quantum kernel machine}

\author{Kouhei~Nakaji}
\affiliation{Quantum Computing Center, Keio University, Hiyoshi 3-14-1, 
Kohoku-ku, Yokohama 223-8522, Japan}
\affiliation{Research Center for Emerging Computing Technologies, National Institute of Advanced Industrial Science and Technology (AIST), 1-1-1 Umezono, Tsukuba, Ibaraki 305-8568, Japan.}
\affiliation{
Current Address: Department of Chemistry, University of Toronto, Toronto, Ontario M5G 1Z8, Canada
}

\author{Hiroyuki~Tezuka}
\affiliation{Sony Group Corporation, 1-7-1 Konan, Minato-ku, Tokyo, 108-0075, Japan}
\affiliation{Quantum Computing Center, Keio University, Hiyoshi 3-14-1, Kohoku-ku, Yokohama 223-8522, Japan}
\affiliation{Graduate School of Science and Technology, Keio University, 
Hiyoshi 3-14-1, Kohoku-ku, Yokohama, 223-8522, Japan}

\author{Naoki~Yamamoto}
\affiliation{Quantum Computing Center, Keio University, Hiyoshi 3-14-1, 
Kohoku-ku, Yokohama 223-8522, Japan}
\affiliation{Department of Applied Physics and Physico-Informatics, 
Keio University, Hiyoshi 3-14-1, Kohoku-ku, Yokohama, 223-8522, Japan}

\date{August 2022}

\begin{abstract}
Quantum machine learning (QML) is the spearhead of quantum computer applications. In particular, quantum neural networks (QNN) are actively studied as the method that works both in near-term quantum computers and fault-tolerant quantum computers. 
Recent studies have shown that supervised machine learning with QNN can be interpreted as the quantum kernel method (QKM), suggesting that enhancing the practicality of the QKM is the key to building near-term applications of QML. 
However, the QKM is also known to have two severe issues. 
One is that the QKM with the (inner-product based) quantum kernel defined in the original large Hilbert space does not generalize; namely, the model fails to find patterns of unseen data. The other one is that the classical computational cost of the QKM increases at least quadratically with the number of data, and therefore, QKM is not scalable with data size. 
This paper aims to provide algorithms free from both of these issues. 
That is, for a class of quantum kernels with generalization capability, we show that the QKM with those quantum kernels can be made scalable by using our proposed deterministic and random features.  
Our numerical experiment, using datasets including $O(1,000) \sim O(10,000)$ training data, supports the validity of our method.
\end{abstract}
\maketitle

\section{Introduction}

Machine learning (ML) algorithms have been rapidly evolving, accompanied by the increase in the volume and complexity of data. 
Demands for more efficient models to describe those data ignited the interest in ML utilizing quantum computing, i.e., quantum machine learning (QML) \cite{biamonte2017quantum,schuld2015introduction,ciliberto2018quantum,lloyd2013quantum}. 
Among many proposals in QML, we focus on the so-called quantum-classical hybrid algorithms that work both in near-term quantum computers and fault-tolerant quantum computers. 
There are two directions in the hybrid algorithms: the quantum neural network (QNN) \cite{farhi2018classification} and the quantum kernel method (QKM) \cite{havlivcek2019supervised,schuld2019quantum}. 
A notable fact is that QKM can find a globally optimal solution better than QNN in supervised machine learning because, even though a globally optimal solution of QKM and infinite depth QNN are equivalent \cite{Schuld2021-tm}, the non-convex optimization in QNN is much more problematic than the convex optimization in QKM due to the problems such as the barren plateau phenomenon \cite{mcclean2018barren}. 
Also, Ref.~\cite{Huang2021-pq} gives a characterization of QKM that has a quantum advantage over classical ML algorithms.

QKM, however, has two serious issues: the generalization capability issue and the scalability issue, both of which are shared with the classical kernel method.
The generalization capability issue states as follows. 
Suppose that a quantum kernel, which represents a similarity between data, is a function of the distance between data-embedded quantum states in a Hilbert space with exponentially large dimensions. 
The original quantum kernel \cite{havlivcek2019supervised}, defined by the inner product of pure quantum states, is one of such kernels. 
However, in this case, the difference of kernel for different data decreases exponentially fast as the size of Hilbert space grows \cite{thanasilp2022exponential}, and the resulting Gram matrix becomes close to the identity matrix \cite{Huang2021-pq,Kubler2021-pm}. 
The QKM with such quantum kernels does not generalize, meaning that it fits with the pattern of the training data, but fails to find patterns of the unseen data.

The scalability issue states that, given the number of training data as $M$, we need at least $O(M^2)$ classical computation, which is problematic when $M$ is large (See  \cite{Schuld2021-tm}). 
For example, in the kernel Ridge regression, $O(M^3)$ computation is required for diagonalizing the Gram matrix. 
This scalability issue is not unique to the quantum case; the issue of poor scalability with data volume exists in the general kernel method and severely limits its applicability. 
Hence, in the classical case, various scalable kernel methods have been proposed \cite{Rahimi2007-om,williams2000using,smola2000sparse,hsieh2014divide,zhang2013divide,liu2020learning}. 
Among them, the most popular framework is the method using {\it random Fourier features} \cite{Rahimi2007-om};
the idea is to approximate the target kernel by the inner products of probabilistically generated feature vectors, where these two quantities are 
guaranteed to be close with each other. 
The solution to the linear machine learning problem using those approximating features achieves a competitive performance to the original kernel method, while the computational complexity to have the approximating solution is much reduced compared to the original one. 
For example, given a dimension of the features as $D$, the computational complexity of training the linear regression model with the features is reduced to be $O(M D^2)$, while the corresponding kernel ridge regression requires $O(M^3)$ computation. 
Note that the random Fourier features require the kernel to be shift-invariant, and thus we cannot directly apply this technique to the quantum kernel, which is generally not shift-invariant. 
Still, the idea of approximating the quantum kernel with an inner product of low-dimensional features is an attractive approach to scalable QKM.

This paper proposes QKM free from the generalization capability issue and the scalability issue. 
We first examine a class of quantum kernels with generalization capability, which we call the quantum kernel with generalization capability (QKGC). 
The quantum kernel with reduced density matrix (RDM) is previously proposed \cite{Huang2021-pq,Kubler2021-pm} as QKGC. As other possible QKGC, we examine a quantum kernel with reduced observable (RO), which we define by a distance between observables. However, our theoretical analysis shows that the quantum kernel with RO is almost equivalent to the quantum kernel with RDM. We also examine naively constructible QKGCs, but it is shown that they lack quantum advantage. 
These observations led us to examine the scalability property of QKGC with RDM.

Our main contribution is to show that the scalability issue can be solved for QKGC with RDM. 
Specifically, we rigorously show that QKGC with RDM can be always approximated by the inner products of low-dimensional features, with a guaranteed error bound. 
Therefore, as with the random Fourier features, we can obtain the approximating solution with those features to the original QKM problem with reduced computational complexity.
Our numerical experiment, using datasets including $O(1,000) \sim O(10,000)$ training data, supports the validity of our proposed method.

The rest of the paper is organized as follows. In Section~\ref{section:background}, we briefly review the background of QKM. The generalization capability issue and the scalability issue of QKM are also discussed in the section. In Section~\ref{section:QKGC}, we discuss QKGC with RDM as a solution for the generalization capability issue. 
Section~\ref{section:feature} is dedicated to describing our method for approximating QKGC. 
Finally, we demonstrate the performance of the proposed QKM method in some benchmark ML problems in Section~\ref{section:numerical}.


\section{Conventional QKM}
\label{section:background}

In this section, we first review the basics of the QKM; 
in particular, the correspondence between QNN and QKM is discussed. 
Next, we show the issues of the QKM; the issue of generalization capability and the issue of computational complexity. 
Finally, we give some remarks on the recently proposed data reuploading model \cite{gil2020input,perez2020data,schuld2021effect} in light of the QKM.

\subsection{Preliminary}
\label{section:generality}

\subsubsection{Basics of QKM}

Let $\mathcal{X}$ be an instance space and $\mathcal{Y}$ be a label space and there is a hidden relationship between $\vec{x} \in \mathcal{X}$ and 
$y \in \mathcal{Y}$ as $y = f^{\ast}(\vec{x})$. 
The goal of supervised ML is to train a model function $f_{\vec{\theta}}:\mathcal{X}\rightarrow\mathcal{Y}$ so that $f_{\vec{\theta}}(\vec{x}) \simeq f^{\ast}(\vec{x})$ by using a training data $\dataspace_{\rm data} = \{\vec{x}_j, y_j \}_{j=1}^M$, where $\vec{\theta}$ is model parameters, $\vec{x}_j \in \mathcal{X}$, and $y_j \in \mathcal{Y}$.

In the QKM, we construct the model function by 
\begin{equation}
\label{eq:model-function}
	f_{\vec{a}}(\vec{x}) = \sum_{j=1}^M a_j Q(\vec{x}_j, \vec{x}), 
\end{equation}
with $Q(\vec{x}, \vec{x}^{\prime})$ as a quantum kernel and $\vec{a} = \{a_j\}_{j=1}^M$ as real parameters to be determined (to clearly distinguish the parameters from those contained in the quantum circuit, we write the parameters of the model function in kernel method as $\vec{a}$ instead of $\vec{\theta}$). 
The quantum kernel is computed from two quantum states in which data is embedded. 
Namely, for a quantum state (density operator) $\rho(\vec{x})$ with $\vec{x}$ a classical data, we compute the quantum kernel as $Q(\vec{x}, \vec{x}^{\prime}) = h(\rho(\vec{x}), \rho(\vec{x}^{\prime}))$, where $h$ is a real-valued function that takes two density operators. 
Note that $\rho(\vec{x})$ can be a ``quantum data" generated by some quantum process, in which case $\vec{x}$ represents a label or parameter for identifying 
such quantum data.

Once the quantum kernels are computed for all training dataset and the Gram matrix is constructed, we follow the standard ML algorithm such as the kernel Ridge regression and the support vector machine \cite{murphy2012machine} to have the model function. 
That is, we train the parameters $\vec{a}$ so that a cost function $L(\vec{a}, \lambda,  \dataspace_{\rm data})$ defined by 
\begin{equation}
\label{eq:cost}
	L(\vec{a}, \lambda, \dataspace_{\rm data}) = 	  \frac{1}{M}\sum_{k=1}^M \ell (y_k, f_{\vec{a}}(\vec{x}_k)) + \lambda \vec{a}^T \vec{Q} \vec{a},
\end{equation}
is minimized, where $\lambda$ is the regularization parameter and  $\vec{Q}$ is the Gram matrix with $(j, k)$ element $Q_{jk} = Q(\vec{x}_j, \vec{x}_k)$. 
The function $\ell: \mathcal{Y} \times \mathcal{Y} \rightarrow \r$ depends on the algorithm; for example, the mean squared loss $\ell(y, f(\vec{x})) = (y - f(\vec{x}))^2/2$ corresponds to the kernel Ridge regression. 
As far as $\ell$ is a convex function, minimization of $L$ with respect to $\vec{a}$ is the convex optimization problem.

The QKM is recognized as a promising variant of the kernel method, because the quantum kernel utilizes exponentially high dimensional feature space of $\rho(\vec{x})$, which is intractable by any classical means but possible via quantum devices with many qubits.

\subsubsection{Correspondence between QKM and QNN}

The correspondence between QKM and QNN is discussed in \cite{Schuld2021-tm}. 
In the QNN models, the model function is constructed as 
\begin{equation}
	f_{\vec{\theta}}(\vec{x}) = {\rm Tr}(
	O U(\vec{\theta})\rho(\vec{x})U^{\dagger}(\vec{\theta})
	)), 
\end{equation}
with a data-embedded quantum state $\rho(\vec{x})$, parameters $\vec{\theta}$, and an observable $O$. 
This model function can be evaluated on a quantum circuit; namely, the data encoding process generates $\rho(\vec{x})$, which then goes through a parameterized circuit $U(\vec{\theta})$, and finally the expectation value of $O$ is computed.

Let $\mathcal{O}$ be a Hermitian observable defined by 
\begin{equation}	
\label{eq:m}
\mathcal{O} = U^{\dagger}(\vec{\theta}) O U(\vec{\theta}).
\end{equation}
Then $f_{\vec{\theta}}(\vec{x}) = f(\mathcal{O}, \vec{x}) = {\rm Tr}(\mathcal{O}\rho(\vec{x}))$. 
Though the matrix elements of $\mathcal{O}$ are constrained such that they have to satisfy Eq.~\eqref{eq:m}, let us relax the constraint and suppose that $\mathcal{O}$ can be an arbitrary Hermitian operator. 
Now consider the cost function 
\begin{equation}	
\label{eq:linear-cost}
L(\mathcal{O}, \lambda, \dataspace_{\rm data}) 
= \frac{1}{M}\sum_{k=1}^M \ell(y_k, f(\mathcal{O}, \vec{x}_k)) + \lambda {\rm Tr}(\mathcal{O}^{\dagger}\mathcal{O}), 
\end{equation}
and suppose that the minimizer of the cost is $\mathcal{O}=\mathcal{O}^{opt}$. The representer theorem states that the optimal function with $\mathcal{O}^{opt}$ can be written as
\begin{equation}
	f(\mathcal{O}^{opt}, \vec{x}) 
	= \sum_{j=1}^{M} a_j^{opt} Q(\vec{x}_j, \vec{x}),
\end{equation} 
where $Q(\vec{x}, \vec{x}^{\prime}) = {\rm Tr}(\rho(\vec{x})\rho(\vec{x}^{\prime}))$. 
The coefficients $\vec{a}^{opt} = \{a_j^{opt}\}_{j=1}^M$ are the minimizer of \eqref{eq:cost} with the same $\ell$ and $\lambda$ as those appearing in \eqref{eq:linear-cost}.

The consequence of the above discussion is that the function obtainable by using the QNN is at most as good as that obtained by using the QKM. 
Also, from another viewpoint, we should use the QKM rather than the QNN; 
because $\mathcal{O}$ may not be able to take $\mathcal{O}^{opt}$ due to the constraint \eqref{eq:m}, and the optimization of the circuit $U(\vec{\theta})$ in the QNN is, in general, much more difficult than the convex optimization of $\vec{a}$ in the QKM. 
However, the QKM has two serious issues, which will be shown in the next subsection.

Before moving forward, note that the above discussion regarding the correspondence between QNN and QKM does not hold for the recently proposed data re-uploading model \cite{gil2020input,perez2020data,schuld2021effect}, which is a variant of QNN. 
We will give some remarks on the correspondence between the data re-uploading QNN model and QKM in Section~\ref{section:reuploading}.

\subsection{Issues of the conventional QKM}
\label{section:issues}


\subsubsection{Issue regarding the generalization capability}

For pure quantum states $\ket{\psi(\vec{x})}$ with $\vec{x} \in \dataspace$, the original quantum kernel \cite{havlivcek2019supervised} is defined as the inner product of two pure quantum states:
\begin{equation}
\label{eq:naive-kernel}
	Q(\vec{x}, \vec{x}^{\prime}) = 
	|\langle\psi(\vec{x})|\psi(\vec{x}^{\prime})\rangle|^2.
\end{equation}
The model function with this kernel, $f_{\vec{a}}(\vec{x})$ given in Eq.~\eqref{eq:model-function}, can learn any function $f^*(\vec{x})$ of the form $f^*(\vec{x}) = \langle \psi(\vec{x})|\mathcal{M}|\psi(\vec{x})\rangle$ with any observable $\mathcal{M}$.

However, the subsequent works \cite{Huang2021-pq,Kubler2021-pm} show that it is practically impossible to train the model function $f_{\vec{a}}(\vec{x})$ with the kernel \eqref{eq:naive-kernel}, if the dimension of Hilbert space of $|\psi(\vec{x})\rangle$ is large. 
More concretely, $M \geq \Omega(2^n)$ of training data is necessary for training $f_{\vec{a}}(\vec{x})$ so that it well approximates $f^{\ast}(\vec{x})$. 
A brief explanation is as follows. 
Given a probability distribution of data as $P_{\rm data}$, let us consider two different samples $\vec{x}$ and $\vec{x}^{\prime}$ sampled from $P_{\rm data}$. 
Then, the probability that $Q(\vec{x}, \vec{x}^{\prime})$ is larger than a certain value $\epsilon$, is exponentially suppressed by the factor ${\rm poly}(1/\epsilon, M, 1/2^n)$, where, again, $M$ is the number of data and we denote a polynomial function of $(a, b, c)$ by ${\rm poly}(a, b, c)$. 
Thus, as far as $M \ll 2^n$, $f_{\vec{a}}({\bf x}_k) \simeq a_k~(\forall k)$ and the coefficients are trained to be $a_k \simeq y_k~(\forall k)$. 
This means that, even though the resulting model function $f_{\vec{a}}(\vec{x})$ well reproduces the label of the training data as $f_{\vec{a}}(\vec{x}_k) \simeq y_k$, it holds $f_{\vec{a}}(\vec{x}) \simeq 0$ for almost all data $\vec{x}$ sampled from $P_{\rm data}$. 
Namely, the function trained with the kernel \eqref{eq:naive-kernel} does not have the generalization capability. 
The same argument is true for the case when two quantum states are mixed states $Q(\vec{x}, \vec{x}^{\prime}) = {\rm Tr}(\rho(\vec{x}) \rho(\vec{x}^{\prime}))$ as far as the dimension of $\rho(\vec{x})$ is exponentially large.


\subsubsection{Issue regarding the computational complexity}

The kernel method, including QKM, generally has the scalability issue; namely, the time complexity and the spatial complexity increase drastically when data volume increases.
Suppose that the kernel function $Q(\vec{x}, \vec{x}^{\prime})$ is defined in the instance space $\vec{x}, \vec{x}^{\prime} \in \mathcal{X}$. 
The poor scalability comes from the computation related to Gram matrix $\vec{Q}$. 
For example, in the kernel Ridge regression, the time complexity scales as $O(M^3)$ (recall that $M$ is the number of training data), since we need to compute the inverse of $\vec{Q}$. 
Also, the spatial complexity scales as $O(M^2)$, since we need to store all components of $\vec{Q}$. 
These scalability issues hinder adopting the QKM to problems with a large dataset.

In the context of the classical kernel method, several solutions have been proposed \cite{Rahimi2007-om,williams2000using}. 
One of the most promising solutions is the random Fourier features \cite{Rahimi2007-om}, which provides a way of mitigating the scalability issue by approximating the kernel function. 
See Appendix~\ref{section:random-fourier-feature} for a brief review of the random Fourier features. 
A notable point is that we can use this technique only when the kernel is of the following shift-invariant form: 
\begin{equation}
\label{eq:shift-invariant}
Q(\vec{x}, \vec{x}^{\prime}) = Q(\vec{x} - \vec{x}^{\prime}). 
\end{equation}
Because in general the quantum kernel does not satisfy this condition, we need other solutions in the quantum case, which is the topic discussed in  Section~\ref{section:feature}.

\subsection{Remark on the data reuploading model}
\label{section:reuploading}

The data reuploading model \cite{gil2020input,perez2020data,schuld2021effect} is recently proposed as a variant of the QNN. 
Unlike the QNN with single data layer, we repeatedly encode data into the quantum state as follows; 
\begin{equation}
\label{eq:data-reuploading}
\begin{split}	
	&\ket{\psi(\vec{\theta}, \vec{x})} \\
	&=U_L(\vec{\theta}_L) U_{L}(\vec{x}) \cdots
	U_{2}(\vec{\theta}_2) U_{2}(\vec{x}) U_{1}(\vec{\theta}_1) U_{1}(\vec{x})|0\rangle^{\otimes n},
\end{split}
\end{equation}
and build the model function as 
	$f_{\vec{\theta}}(\vec{x}) = \bra{\psi(\vec{\theta}, \vec{x})}O\ket{\psi(\vec{\theta}, \vec{x})}$ in the data reuploading model. 
The repeated encoding increases the number of Fourier components in the model function $f_{\vec{\theta}}(\vec{x})$ and accordingly the expressibility of $f_{\vec{\theta}}(\vec{x})$, as we show in Appendix~\ref{section:merit-data-reuploading}.

The theoretical correspondence between the data reuploading model and the QKM can be seen by embedding trainable parameters in the quantum kernel. 
More concretely, given a quantum state 
\begin{equation}
\begin{split}	
	&\ket{\psi(\tilde{\vec{\theta}}, \vec{x})} \\
	&= U_{L}(\vec{x}) U_{L-1}(\vec{\theta}_{L-1}) \cdots
	U_{2}(\vec{\theta}_2) U_{2}(\vec{x}) U_{1}(\vec{\theta}_1) U_{1}(\vec{x})|0\rangle^{\otimes n},
\end{split}
\end{equation}
 with $\tilde{\vec{\theta}} = \{\vec{\theta}_1, \vec{\theta}_2, \cdots,  \vec{\theta}_{L-1}\}$, let us define the parameter embedded kernel function by 
$Q(\vec{x}, \vec{x}^{\prime}, \tilde{\vec{\theta}}) = {\rm Tr(\rho(\vec{x}, \tilde{\vec{\theta}}) \rho(\vec{x}^{\prime}, \tilde{\vec{\theta}}))}$, where $\rho(\vec{x}, \tilde{\vec{\theta}}) = 	\ket{\psi(\tilde{\vec{\theta}}, \vec{x})}	\bra{\psi(\tilde{\vec{\theta}}, \vec{x})}$. 
Then, the training of $f_{\vec{\theta}}(\vec{x})$ with a certain cost function corresponds to the QKM with the quantum kernel $Q(\vec{x}, \vec{x}^{\prime}, \tilde{\vec{\theta}})$, where parameters $\tilde{\vec{\theta}}$ are fixed. 
Of course, the training of the data reuploading model also includes the training of the parameters $\tilde{\vec{\theta}}$, and therefore, to fully relate the QKM with the data reuploading model, we need to train $\tilde{\vec{\theta}}$ within the quantum kernel framework.

As far as we know, there are few pieces of research dedicated to the training of the quantum kernel. 
For instance, see \cite{wu2021provable} that contains the training of the quantum kernel in the context of quantum phase learning. 
This is mainly due to the issue of scalability discussed above; that is, whenever we iteratively update parameters $\tilde{\vec{\theta}}$, we need to recompute the Gram matrix, which requires $O(M^2)$ computation in each iteration. 
Conversely, if we resolve the scalability issue, we may be able to train the quantum kernel. 
In Section~\ref{section:feature}, we will propose a scalable QKM and thereby discuss a method of training quantum kernel.


\section{Quantum kernels with generalization capability (QKGC)}
\label{section:QKGC}

As shown in Section~\ref{section:issues}, the QKM with the naive quantum kernel \eqref{eq:naive-kernel} lacks the generalization capability. 
In this section, we examine quantum kernels that have the generalization capability of QKM; we simply call such kernels as quantum kernels with the generalization capability (QKGC). 
In Section~\ref{section:QKGC-rdm}, we give a generalized version of QKGC using reduced density matrices (RDMs), which was originally introduced in \cite{Huang2021-pq}. 
We then examine QKGC with reduced observables (ROs) in Section~\ref{section:QKGC-RO}. 
Finally, Section~\ref{section:QKGC-others} is devoted to discussing the other possibility of QKGC without RDM.

\subsection{QKGC with the reduced density matrix (RDM)}
\label{section:QKGC-rdm}

As shown in Section~\ref{section:issues}, the low generalization capability of the QKM attributes to the large dimension of the Hilbert space where $\rho(\vec{x})$ is located. 
Therefore, using a reduced density matrix (RDM) 
$\rhoq{\vec{x}} = {\rm Tr}_{\neq \vec{q}} (\Phi(\rho(\vec{x})))$ instead of  $\rho(\vec{x})$, is a natural solution for the issue, where ${\rm Tr}_{\neq \vec{q}}$ means tracing out the qubits {\it not} included in an integer vector $\vec{q}$ and $\Phi$ is a completely positive and trace-preserving (CPTP) map. 
For restoring the generalization capability, the dimension of Hilbert space where $\rhoq{\vec{x}}$ exists must be smaller than the number of data $M$ \cite{Kubler2021-pm}. 
In the following, we introduce the inner product kernel and the distance kernel defined with $\rhoq{\vec{x}}$, which are the generalized version of QKGC proposed in \cite{Huang2021-pq}.

\subsubsection{Inner product kernel with RDM}
\label{section:inner-product-rdm}


\begin{definition}
Let $\rhoqk{\vec{x}}$ be the $k$-th data encoded RDM. 
The index $k$ runs from $1$ to $K$ with some positive integer $K$. 
The inner product kernel $Q$ is a function $Q: \mathcal{M}\times \mathcal{M} \rightarrow \r$
written as
\begin{align}	
\label{eq:projected-quantum-kernel}
Q(\vec{x}, \vec{x}^{\prime}) &= \sum_{k=1}^{K} \alpha_k {\rm Tr}(\rhoqk{\vec{x}}\rhoqk{\vec{x}^{\prime}}), 
\end{align}
with $\{\alpha_k\}_{k=1}^K$ as positive coefficients.
\end{definition}
For the generalization capability, the dimension of $\rhoqk{\vec{x}}$ is smaller than the number of data. 
The inner product kernel with RDM is a natural extension of the original quantum kernel \eqref{eq:naive-kernel}.

\subsubsection{Distance kernel with RDM}

\begin{definition}
Let $\rhoqk{\vec{x}}$ be the $k$-th data encoded density matrix. 
The distance kernel $Q$ is a function $Q: \mathcal{M}\times \mathcal{M} \rightarrow \r$ written as
\begin{equation}
\label{eq:gaussian}
	Q(\vec{x}, \vec{x}^{\prime}) 
	= 
	f\left( \sum_{k=1}^K \alpha_k \dhsk \right),
\end{equation}
where $f: {\bf R} \rightarrow {\bf R}$ is a function that satisfies $f(0) = 1$, and $d^{\rm HS}$ is the distance $(0 \leq d^{\rm HS} \leq 1)$ defined by 
\begin{equation}
\label{HS dis def}
\dhs = \left\| \frac{\rhoq{\vec{x}} - \rhoq{\vec{x}^{\prime}}}{2} \right\|_{HS},
\end{equation}
with $\|\cdot\|_{HS}$ the Hilbert Schmidt distance.
\end{definition}
In \cite{Huang2021-pq}, the Gaussian kernel, where $K=1$, $a_1 = -\gamma$, and $f(c) = \exp(-c)$, is proposed. The quantum kernel defined in \eqref{eq:gaussian} is a natural extension of it.

\subsection{QKGC with reduced observables (RO)}
\label{section:QKGC-RO}

The above two definitions of quantum kernels are based on the similarity of two data-embedded quantum states.
Another way of building a quantum kernel would be one using two data-embedded observables: 
\begin{equation}	
\begin{split}	
	&Q^{\rm obs}(\vec{x}, \vec{x}^{\prime}) \\
	&\equiv \left|\langle \psi| A(\vec{x}) A(\vec{\vec{x}^{\prime}}) |\psi \rangle \right|^2 \\ 
		&= {\rm Tr}\left[
		\left(A(\vec{x}) \rho A(\vec{x})\right)
		\left(A(\vec{x}^{\prime})
		\rho A(\vec{x}^{\prime})
	\right)
	\right],
\end{split}
\end{equation}
where $\rho = |\psi \rangle\langle \psi|$ is a quantum state. 
Here we define $A(\vec{x})$ by
\begin{equation}
	A(\vec{x}) = \sum_{j} \gamma_j U(\vec{x}) P_j U(\vec{x})^{\dagger}, 
\end{equation} 
where $\{P_j\}$ are non-identity generalized Pauli observables and $\{\gamma_j\}$ are real coefficients. As in the case of the original quantum kernel \cite{havlivcek2019supervised}, if the dimension of Hilbert space where $A(\vec{x})$ is located is exponentially large, all the off-diagonal terms of the Gram matrix generated from the kernel $Q^{\rm obs}(\vec{x}, \vec{x}^{\prime})$ takes an exponentially small value, and hence QKM with the kernel $Q^{\rm obs}(\vec{x}, \vec{x}^{\prime})$ does not have the generalization capability.

We can restore the generalization capability by defining the quantum kernel using the reduced observable (RO) as follows; 
\begin{equation}
\label{eq:tilde-kernel}
	Q^{\rm RO}(\vec{x}, \vec{x}^{\prime}) \equiv \sum_{k=1}^K \alpha_k {\rm Tr}\left(
		\tilde{A}_{\rho}^{\vec{q}_k}(\vec{x}) \tilde{A}_{\rho}^{\vec{q}_k}(\vec{x}^{\prime})
	\right),
\end{equation}
where RO is defined as 
\begin{align}	
	\tilde{A}_{\rho}^{\vec{q}}(\vec{x}) &=
	{\rm Tr}_{\neq \vec{q}}\left(
	A(\vec{x}) \rho A(\vec{x}) 
	\right). 
\end{align}
Note that this can be represented as 
\begin{equation}
\label{eq:sum-reduced-obs}
\begin{split}	
	&\tilde{A}_{\rho}^{\vec{q}}(\vec{x}) \\
	&= \sum_{j,\ell} \gamma_j \gamma_{\ell}
	{\rm Tr}_{\neq \vec{q}} \left(
	U(\vec{x})P_j U(\vec{x})^{\dagger} \rho U(\vec{x})P_{\ell} U(\vec{x})^{\dagger}) 
	\right). 
\end{split}
\end{equation}
As in the case of QKGC with RDM, we assume that the dimension of Hilbert space after tracing out is smaller than the number of data.

The above-defined QKGC with RO \eqref{eq:tilde-kernel} seems to be an alternative to QKGC with RDM. However, QKGC with RO is equivalent to QKGC with RDM other than exponentially suppressed terms. 
Actually, in Appendix~\ref{section:analysis-ro}, we show that the terms with $j\neq \ell$ in \eqref{eq:sum-reduced-obs} are exponentially suppressed for almost all of $U(\vec{x})$. Therefore, the following quantum kernel
\begin{equation}
\label{eq:qkgc-ro}
		\tilde{Q}^{\rm RO}(\vec{x}, \vec{x}^{\prime}) \equiv  \sum_{k=1}^K \alpha_k {\rm Tr}\left(
{A}_{\rho}^{\vec{q}_k}(\vec{x}) {A}_{\rho}^{\vec{q}_k}(\vec{x}^{\prime})
	\right),
\end{equation}
where
\begin{equation}
	A^{\vec{q}}_{\rho}(\vec{x}) = \sum_{j} \gamma_j^2 {\rm Tr}_{\neq \vec{q}}
	\left(
	U(\vec{x})P_j U(\vec{x})^{\dagger} \rho U(\vec{x})P_{j} U(\vec{x})^{\dagger} \right), 
\end{equation}
is equal to \eqref{eq:tilde-kernel} other than the exponentially suppressed terms for almost all of $U(\vec{x})$. 
Since $P_j$ is a unitary operator, 
\begin{equation}
	\rho_j(\vec{x}) = U(\vec{x})P_j U(\vec{x})^{\dagger}\rho U(\vec{x})P_{j} U(\vec{x})^{\dagger} 
\end{equation}
is a density operator and ${\rm Tr}_{\neq \vec{q}}(\rho_j(\vec{x}))$ is a RDM. 
Thus, \eqref{eq:qkgc-ro} can be written in the form of the inner product kernel with RDM discussed in Section~\ref{section:QKGC-rdm}.

\subsection{Other candidates for QKGC}
\label{section:QKGC-others}

Recall that the root cause of the low generalization capability is that the kernel $Q(\vec{x}, \vec{x}^{\prime})$ becomes almost zero if $\vec{x}$ and $\vec{x}^{\prime}$ are different, and the Gram matrix $\vec{Q}$ is almost diagonal. 
Thus, all we need to restore the generalization capability is to modify $Q(\vec{x}, \vec{x}^{\prime})$ so that it can take a large value even if $\vec{x}$ and $\vec{x}^{\prime}$ are different. 
Actually, the quantum kernels with RDM discussed in Section~\ref{section:QKGC-rdm} satisfy this condition.

Therefore it is interesting to ask if we could construct a good quantum kernel without using RDM or RO. 
In Appendix~\ref{section:QKGC-others-example}, we examine two types of QKGC  without using RDM; the quantum kernel with tiny coefficients and the quantum kernel with a few Fourier components. 
The result is that, even though both of these quantum kernels may have generalization capability, they have no quantum advantages.

Let us summarize Section~\ref{section:QKGC}. 
QKGC with RDM successfully restores the generalization capability. 
QKGC with RO we proposed, is another possible QKGC, but it is almost equivalent to QKGC with RDM. 
Moreover, we examined some other types of QKGC, but they lack quantum advantage.
Based on this fact, in the rest of this paper, we will focus on QKGC with RDM as the quantum kernel that enjoys both the generalization capability and, further, may have
the quantum advantage.


\section{Feature representation of QKGC}
\label{section:feature}

The goal of this section is to show that QKGC with RDM has scalability properties, in addition to the generalization capability. 
More concretely, we show that for any QKGC with RDM, $Q(\vec{x}, \vec{x}^{\prime})$, there always exists a feature vector $\vec{z}(\vec{x})$ that satisfies
\begin{equation}
	Q(\vec{x}, \vec{x}^{\prime}) \simeq \vec{z}(\vec{x})^T \vec{z}(\vec{x}^{\prime}),
\end{equation}
where $T$ denotes the transpose of the vector.

It is well known that the scalability issue is much-alleviated by using the feature representation $\vec{z}(\vec{x})$ instead of the quantum kernel, as long as we take the dimension of the vector $\vec{z}(\vec{x})$ smaller than the number of data $M$. 
For example, in the kernel Ridge regression problem, if the kernel can be approximated as the inner product of feature vectors, then the problem boils down to that of a simple linear regression problem, and as a result, the computational complexity improves from $O(M^3)$ to $O(MD^2)$ and the spatial complexity improves from $O(M^2)$ to $O(MD)$.

In Section~\ref{section:feature-representation-inner}, we identify the feature vector for the inner product kernel with RDM. 
Then the feature representation of the distance kernel with RDM is given in Section~\ref{section:feature-representation-shift}. 
In Section~\ref{section:summary-discussion-features}, we summarize those features and discuss their advantages.

\subsection{Feature representation of the inner product kernel with RDM}
\label{section:feature-representation-inner}

Here, we show that the summand of the right-hand side of the inner product kernel \eqref{eq:projected-quantum-kernel} can be represented as 
\begin{equation}
{\rm Tr}(\rhoqk{\vec{x}}\rhoqk{\vec{x}^{\prime}}) = \vec{z}(k, \vec{x})^T \vec{z}(k, \vec{x}), 
\end{equation}
with a feature vector $\vec{z}(k, \vec{x})$. 
Then, it directly holds that 
\begin{align}
Q(\vec{x}, \vec{x}^{\prime}) &= \sum_{k=1}^{K} \alpha_k {\rm Tr}(\rhoqk{\vec{x}} \rhoqk{\vec{x}^{\prime}}) \\
&= \vec{z}(\vec{x})^T \vec{z}(\vec{x}^{\prime}), 
\end{align}
with
\begin{equation}
\vec{z}(\vec{x})
 = \left(
\begin{array}{c}
 \sqrt{\alpha_1}\vec{z}(1, \vec{x}) \\
 \sqrt{\alpha_2} \vec{z}(2, \vec{x}) \\
 \vdots \\
  \sqrt{\alpha_k}\vec{z}(K, \vec{x})
\end{array}
\right).
\end{equation}
If $K$ is large, instead of concatenating all $\{\sqrt{\alpha_k} \vec{z}(k, \vec{x})\}_{k=1}^K$, we may sample the set of indices $\{k_1, k_2, \cdots, k_m \}$ $(m\ll K)$ according to the probability distribution $p(k) = \sqrt{\alpha_k}/\sum_{k=1}^K \sqrt{\alpha_k}$ and build $\vec{z}(\vec{x})$ as 
\begin{equation}
	\vec{z}(\vec{x})  
	= \left(
\begin{array}{c}
\vec{z}(k_1, \vec{x}) \\
\vec{z}(k_2, \vec{x}) \\
 \vdots \\
 \vec{z}(k_m, \vec{x})
\end{array}
\right).
\end{equation}

For simplifying the notation, in what follows, we consider the quantum kernel 
of the form $Q(\vec{x}, \vec{x}^{\prime}) = {\rm Tr}(\rhoq{\vec{x}} \rhoq{\vec{x}^{\prime}})$ and will show that $Q(\vec{x}, \vec{x}^{\prime})$ can be represented by an inner product of two feature vectors. 
The following theorem is the key to constructing the feature vector.

\begin{th.}
\label{th:quantum-theorem}	
Let $n_{\vec{q}}$ be the number of qubits in $\rho_{\vec{q}}(\vec{x})$ and $\Nq=2^{n_\vec{q}}$. 
Then the following equality holds:
\begin{equation}
\begin{split}	
\label{eq:quantum-theorem}	
	&{\rm Tr}(\rhoq{\vec{x}}\rhoq{\vec{x}^{\prime}}) \\
	&= \frac{1}{\Nq}\sum_{j=1}^{\Nq^2} {\rm Tr}\left(\pqji \Phi\left(\rho({\bf x})\right) \right)  \\
	&~~~~~~~~~~~~~{\rm Tr}\left(\pqji \Phi\left(\rho({\bf x}^{\prime})\right) \right),	
\end{split}
\end{equation}
where $\pqj \in \{I, X, Y, Z\}^{\otimes n_{\vec{q}}}$ are operators acting on 
the qubits specified by $\vec{q}$ and 
$\iq$ is the identity operator acting on all the other qubits. 
\end{th.}

The proof is given in Appendix~\ref{section:proof-theorem-projected}.
From Theorem~\ref{th:quantum-theorem}, we can represent the kernel 
$Q(\vec{x}, \vec{x}^{\prime})={\rm Tr}(\rhoq{\vec{x}}\rhoq{\vec{x}^{\prime}})$ using two vectors as 
\begin{equation}
\label{Th 1 corollary}
   Q(\vec{x}, \vec{x}^{\prime}) 
   = \vec{z}(\vec{x})^T \vec{z}(\vec{x}^{\prime}), 
\end{equation}
where $\vec{z}(\vec{x}) = \{z_j(\vec{x})\}_{j=1}^{\Nq^2}$ with 
\begin{equation}
z_j(\vec{x}) = \frac{ {\rm Tr}\left(\pqji \Phi\left(\rho({\bf x}^{\prime})\right) \right)}{\sqrt{\Nq}}. 
\end{equation}
Note that the vector $\vec{z}(\vec{x})$ can be efficiently computed using a quantum device. 
We call $\vec{z}(\vec{x})$ the {\it deterministic quantum feature (DQF)}.

We can also estimate the effect of the shot noise. 
Let $\ns$ be the number of measurements to estimate ${\rm Tr}(\pqji \Phi(\rho(\vec{x})))$ for each $j$. 
Then, the probability that the kernel \eqref{Th 1 corollary} is distorted with magnitude $\epsilon>0$ by shot noise is bounded as
\begin{equation*}
	{\rm Prob}\left[{\rm sup}_{\vec{x},\vec{x}^\prime \in \dataspace_{\rm data}}
	\left|\vec{z}(\vec{x})^T \vec{z}(\vec{x}^{\prime}) 
	   - Q(\vec{x}, \vec{x}^{\prime}) \right| \geq \epsilon\right] 
	\leq \delta,
\end{equation*}
if the number of measurements $\ns$ is bounded as
\begin{equation*}
\ns \geq \frac{18\Nq^2}{\epsilon^2} \log\left(\frac{2 M}{\delta}\right).
\end{equation*}
The proof of this fact is given in Appendix \ref{section:shot-noise}.

\subsection{Feature representation of the distance kernel with RDM}
\label{section:feature-representation-shift}

As in the case of the inner product kernel, we focus on the simplified version of the kernel \eqref{eq:gaussian} and show that 
\begin{equation}
	f\left( \dhs \right) 
	\simeq \vec{z}(\vec{x})^T \vec{z}(\vec{x}),
\end{equation}
with a feature vector $\vec{z}(\vec{x})$. 
Recall that $d^{\rm HS}$ is defined using the Hilbert Schmidt distance in Eq.~\eqref{HS dis def}. 
The following theorem is the key to constructing the features. 

\begin{th.}
\label{th:hs-equality}
It holds 
\begin{equation}	
\label{eq:hs-equality}
\begin{split}	
	&f\left( \dhs \right) \\ 
	&= 
{\bf E}_{\vec{\omega} \sim P(\vec{\omega})}
\left[
\cos\left({\rm Tr}(
\vec{\omega}^T \vec{c}(\vec{x})
)
\right) 
\cos\left({\rm Tr}(\vec{\omega}^T \vec{c}(\vec{x}^{\prime}))\right) \right.
\\
&~~\left. 
\sin\left(
{\rm Tr}(
\vec{\omega}^T \vec{c}(\vec{x})
)
\right)
\sin\left(
{\rm Tr}(
\vec{\omega}^T \vec{c}(\vec{x}^{\prime})
)
\right)
\right],
\end{split}
\end{equation}
where $P(\vec{\omega})$ is a probability distribution dependent on $f$, $\vec{\omega} \sim P(\vec{\omega})$ states that $\vec{\omega} \in {\bf R}^{\Nq^2}$ is sampled from 
$P(\vec{\omega})$, and $\vec{c}(\vec{x}) \in {\bf R}^{\Nq^2}$ is a vector with $j$-th element defined by $c(\vec{x})_{j} = {\rm Tr}\left(\pqji  \Phi\left(\rho({\bf x})\right) \right)/\Nq$.
\end{th.}

We give the proof in Appendix~\ref{section:proof-theorem-distance}.
As the corollary of Theorem~\ref{th:hs-equality}, we can construct a feature vector corresponding to the distance kernel. Let $z_{\vec{\omega}}^{s}(\vec{x}) = \sin({\rm Tr}(\vec{\omega}^T \vec{c}(\vec{x})))$ and $z_{\vec{\omega}}^c(\vec{x}) = \cos({\rm Tr}(\vec{\omega}^T \vec{c}(\vec{x})))$. Sampling $\vec{\omega}$ from $P(\vec{\omega})$, we can generate an array of $\vec{\omega}$ as $\{\vec{\omega}^{(1)}, \vec{\omega}^{(2)}, \cdots, \vec{\omega}^{(D/2)} \}$. 
Then, defining a feature vector $\vec{z}(\vec{x})$ by 
\begin{equation}
	\vec{z}(\vec{x}) = 
	\left(
	\begin{array}{c}
		\vec{z}^s(\vec{x})\\
		\vec{z}^c(\vec{x})
	\end{array}\right),
\end{equation}
where
\begin{eqnarray}
\hspace{-1cm}
	\vec{z}^s(\vec{x}) &= \sqrt{\frac{2}{D}}
	\left(
	\begin{array}{c}
		z_{\vec{\omega}_1}^{s}(\vec{x}) \\ 
		z_{\vec{\omega}_2}^{s}(\vec{x}) \\ 
		\vdots \\
		z_{\vec{\omega}_{D/2}}^{s}(\vec{x}) \\ 
	\end{array}
	\right), 
\\
	\vec{z}^c(\vec{x}) &= \sqrt{\frac{2}{D}}
	\left(
	\begin{array}{c}
		z_{\vec{\omega}_1}^{c}(\vec{x}) \\ 
		z_{\vec{\omega}_2}^{c}(\vec{x}) \\ 
		\vdots \\
		z_{\vec{\omega}_{D/2}}^{c}(\vec{x}) \\ 
	\end{array}
	\right), 
\end{eqnarray}
we obtain $f\left( \dhs \right)\simeq \vec{z}(\vec{x})^T \vec{z}(\vec{x}^{\prime})$. 
We call this $\vec{z}(\vec{x})$ as the {\it random quantum feature (RQF)}.

There are two sources of kernel approximation error: the error due to the shot noise and the error from sampling due to the limited number of samples $D$. Combining the effects of those errors, we obtain the following theorem.

\begin{th.}
\label{th:hs-error}
Let $L_f$ be the Lipschitz constant of f with respect to the $L_1$ norm. 
Given $Q(\vec{x}, \vec{x}^{\prime}) = f\left(\dhs \right)$, the following bound holds for kernel approximation error:
\begin{equation}	
\label{eq:hs-error}
\begin{split}	
	 {\rm Prob}\left[{\rm sup}_{\vec{x},\vec{x}^\prime \in \dataspace_{\rm data}}
	 \left|\vec{z}(\vec{x})^T\vec{z}(\vec{x}^{\prime}) - Q(\vec{x}, \vec{x}^{\prime})\right| \geq \epsilon\right] \leq \delta, 
\end{split}
\end{equation}
if 
	\begin{equation}
		\ns \geq \frac{8\Nq^2 L_f^2}{\epsilon^2} \log\left(\frac{4 M}{\delta}\right),
	\end{equation} 
and 
	\begin{equation}
		D \geq \frac{32(\Nq^2 + 2)}{\epsilon^2} \log\left(\frac{2112\sigma_p^2}{\epsilon^2 \delta}\right), 
	\end{equation}
where $\sigma_p = \expect{\vec{\omega}\sim P(\vec{\omega})}{\vec{\omega}^T \vec{\omega}}$.
\end{th.}

We give the proof in Appendix~\ref{section:proof-distance-error}. 
Therefore, there are strict lower bounds on the number of shots, $\ns$, and the dimension of the feature vector, $D$, to guarantee a given approximation error.

\subsection{Summary and discussion on the use of DQF and RQF}
\label{section:summary-discussion-features}

In Sections~\ref{section:feature-representation-inner} and \ref{section:feature-representation-shift}, we obtained DQF and RQF for two types of QKGC. 
In particular, the dimension of the feature vector, $D$, needs to be $\Nq^2$ in the case of DQF and $O(\Nq^2/\epsilon^2)$ in the case of RQF, where $\epsilon$ is a given kernel approximation error.

\begin{figure}
\includegraphics[width=0.5\textwidth]{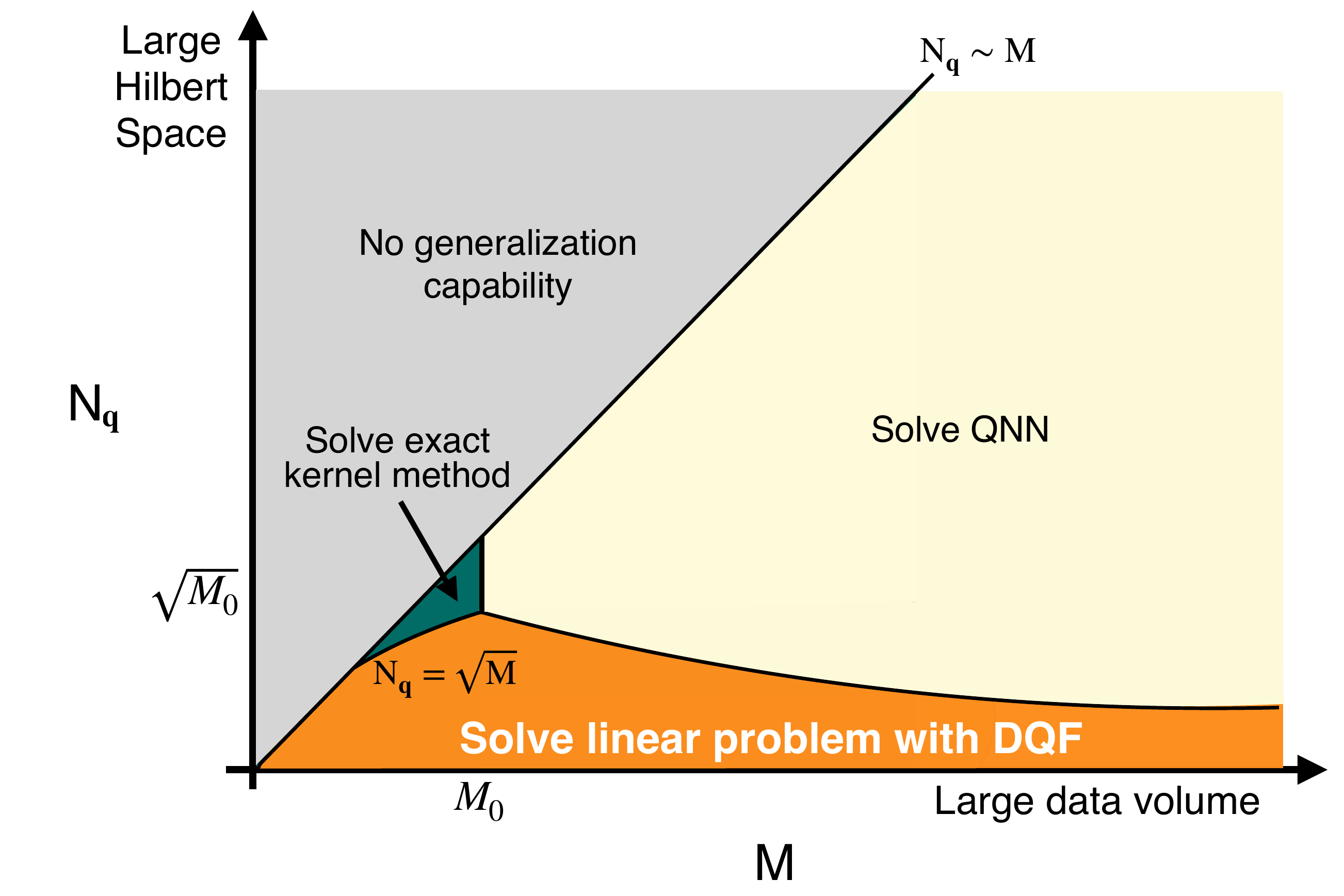}	
\caption{Regions where DQF has an advantage over the conventional quantum kernel.  
The horizontal axis denotes the number of data $M$, and the vertical axis denotes the dimension of the Hilbert space. 
}
\label{fig:contour}
\end{figure}

In Fig.~\ref{fig:contour}, we plot the two-dimensional map with respect to the number of data $M$ and the dimension of the reduced Hilbert space $\Nq$, and specify the region such that solving the linear regression problems with DQF has an advantage over the quantum kernel method with the conventional quantum kernel \eqref{eq:naive-kernel}. For simplicity, we assume that the training cost for the kernel method is $O(M^3)$ and that for the corresponding linear model with DQF is $O(M D^2)$ with $D=N_{\vec{q}}^2$ again, which are the cases in the kernel ridge regression. However, a similar figure can be drawn in the case of another kernel method, such as the support vector machine. 
In the region where $\Nq \gtrsim M$, the QKM with DQF does not have the generalization capability because the discussion in Section II B 1 is applied to this case as well; the corresponding region is colored gray. 
When $M$ is small enough ($M \leq M_0$ with $M_0$ as the maximum data size where $O(M_0^3)$ classical computation is executable), and $\Nq$ is between $\sqrt{M}$ and $M$, we should use the quantum kernel method with the exact quantum kernel because $ O(M D^2) \gtrsim O(M^3)$ in the region; the region is colored with green. 
When $\Nq$ is small enough to satisfy $\Nq < \sqrt{M}$ and $O(MD^2)$ classical computation is executable, i.e. $O(MD^2) < O(M_0^3)$, 
we should use the linear model with DQF in the quantum kernel method.  
The region is colored orange. 
Otherwise, we can use neither the quantum kernel method nor the linear model with DQF, and we may need to use a QNN; the region is colored yellow. 
A similar contour can be drawn in the case of RQF.


Note that in the region where $\Nq < \sqrt{M}$ and $M < M_0$, both the quantum kernel method with the exact quantum kernel and the linear model with DQF are usable. 
However, our view is that it is better to use the linear model with DQF in this region because of the flexibility of DQF. 
In particular, we can extend the QKM to the one including the training of the quantum kernel when using DQF. 
More specifically, by embedding the parameters $\vec{\theta}$ in RDM, we obtain each element of DQF as
\begin{equation}
z_j(\vec{x}, \vec{\theta}) = \frac{ {\rm Tr}\left(\pqji \Phi\left(\rho({\bf x}, \vec{\theta})\right) \right)}{\sqrt{\Nq}}. 
\end{equation}
Then, given the linear model using the features as
\begin{equation}
	f(\vec{x}, \vec{\theta}) = \vec{w}^T \vec{z}(\vec{x}, \vec{\theta})
\end{equation}
and the cost function as 
\begin{equation}
	\mathcal{L} = \sum_{m=1}^M C(f(\vec{x}_m, \vec{\theta}), y_m), 
\end{equation}
we can minimize $\mathcal{L}$ not only for $\vec{w}$ but for $\vec{\theta}$. 
The same arguments hold for the case of RQF. 
Actually, we can efficiently compute the gradient of the cost with respect to $\vec{\theta}$ using the parameter-shift rule \cite{Crooks2019GradientsOP}, and therefore, we can update $\vec{\theta}$ by the gradient descent. 
Note that training the parameters $\vec{\theta}$ corresponds to learning the quantum kernel. 
As noted in Section~\ref{section:reuploading}, QKM involving such kernel learning process corresponds to the data reuploading model. 
Certainly, the optimization of $\vec{\theta}$ is a non-convex problem, and the difficulty of training QNNs may appear in this problem as well. 
Nevertheless, in our view, the quantum kernel method may still work with even roughly chosen $\vec{\theta}$, as such a difficult optimization problem needs not to be perfectly solved. 
We will study the learning problem of the quantum kernel in a separate work.


\section{Numerical studies}
\label{section:numerical}

In this section, we demonstrate the performance of DQF and RQF introduced in the previous section.  
We perform the regression and the classification tasks using the datasets: `wine quality' and `magic04' shown in TABLE~\ref{table:data}. We encode data so that the number of qubits in $\rho(\vec{x})$ is the same as the dimension of the dataset; for example, in the case of the wine quality dataset, we use 11-qubits. We set the CPTP map $\Phi$ as the identity map. Each kernel is generated by reducing some qubits in $\rho(\vec{x})$. The detail of the data encoding method to build $\rho(\vec{x})$ is shown in Appendix~\ref{section:detail-experiment}. Since the number of training data is large, it is costly to run a kernel method without using DQF or RQF. 
For example, in the case of the wine quality dataset, the number of training data is 4,000, and therefore, to solve the kernel ridge regression, we need $4000^3 \sim 10^{11}$ computation to diagonalize the Gram matrix. Thus, we need a kernel approximation method to mitigate the computational complexity. As the approximation method, we compare DQF/RQF with the Nystrom method \cite{williams2000using}. We review the Nystrom method in Appendix~\ref{section:detail-experiment}.

For the regression task, we solve the linear regression with the regularization parameter $\lambda = 0.001$, where the cost function is set to be the mean squared error. For the classification task, we solve the soft-margin support vector machine with the regularization parameter $C = 1$, where the cost function is set to be the Hinge loss. To perform the support vector machine, we use the package scikit-learn \cite{scikit-learn}.

In addition to checking the performance in the machine learning tasks, we will see how well a quantum kernel is approximated. As the metric to measure the discrepancy between the exact quantum kernel $Q$ and the kernel $\tilde{Q}$ that is computed as the inner products of  DQF or RQF, we use the following distance
\begin{equation}	
d_Q (\tilde{Q}, Q) \equiv \frac{1}{M^2}\sum_{j,k=1}^M \left|\tilde{Q}(\vec{x}_j, \vec{x}_k) - Q(\vec{x}_j, \vec{x}_k)\right|.
\end{equation}
Note that the source of the discrepancy between $\tilde{Q}$ and $Q$ is the shot noise in DQF, while, in the case of RQF, the discrepancy arises from the sampling of the features in addition to the shot noise. 
 Note that, to exactly compute $Q(\vec{x}, \vec{x}^{\prime})$, we use the state vector representation.
Throughout our numerical experiment, we use Qulacs \cite{suzuki2021qulacs} for simulating the quantum processes.

\begin{table*}[ht]
\caption{Details of the dataset used in our numerical experiment. The wine quality dataset is available from \protect \url{https://archive.ics.uci.edu/ml/datasets/wine+quality} and the magic04 dataset is available from \protect \url{https://networkrepository.com/magic04.php}.}
\label{table:data}
\begin{tabular}{p{0.15\textwidth} p{0.25\textwidth} p{0.20\textwidth} p{0.20\textwidth} p{0.20\textwidth}}
\hline\hline
Type & Dataset & Dimension  & \#Training & \#Test\\
\hline
regression & wine quality& 11    &   4,000    &  898            \\

classification & magic04 &   10    & 18,500    &   520    \\
\hline       
\end{tabular}
\end{table*}

\begin{table*}[ht]
\caption{MSE/Accuracy and the Kernel approximation error $d_Q$ in the regression problem with wine quality dataset in the inner product kernel.}
\label{table:performance-inner-product}
\begin{tabular}{p{0.15\textwidth} p{0.25\textwidth} p{0.20\textwidth} p{0.20\textwidth} p{0.20\textwidth}}
\hline\hline
Dataset & Algorithm & \#Features  & $d_Q$ & MSE/Accuracy \\
\hline
\multirow{3}{*}{wine quality} &{DQF}   & 16    &  ${0.007 \pm 0.000}$     &${0.536\pm 0.001}$               \\
&{Nystrom} &   16    & $0.980\pm 0.547$    & $1.393 \pm 0.055$        \\
&{Nystrom}  &  100     &   $0.484\pm 0.139$     &      $0.678 \pm 0.008$ \\
\hline
\multirow{3}{*}{magic04} &DQF   & 64    &     ${0.013 \pm  0.000}$    &  ${82.3 \pm  0.0\%}$           \\
&{Nystrom} &   64    & $11.190\pm 5.823$     & $79.6 \pm  0.0\%$       \\
&{Nystrom}  &  100     & $0.637\pm0.084$  & $79.8 \pm  0.2\%$         \\
\hline       
\end{tabular}
\end{table*}

\subsection*{Experiment using DQF}

In this experiment, we use the following inner product kernel,
\begin{equation}
\label{eq:inner-product-experiment}
	Q(\vec{x}, \vec{x}^{\prime}) =  {\rm Tr}(\rhoq{\vec{x}}\rhoq{\vec{x}^{\prime}}), 
\end{equation}
where $\vec{q}$ is set to be $\vec{q}=\{9,10\}$ in the experiment with the wine quality dataset and $\vec{q} = \{8,9,10\}$ in the experiment with the magic04 experiment. The corresponding feature vector for the quantum kernel \eqref{eq:inner-product-experiment} is DQF introduced in Section~\ref{section:feature-representation-inner}. The number of measurements for building each element of the feature vector is set to be $\ns=500$.

For comparison, we also perform an experiment with another kernel approximation technique called the Nystrom method \cite{williams2000using}. As shown in Appendix~\ref{section:detail-experiment}, the Nystrom method also builds features to approximate the kernel. Note that Nystrom method requires the evaluation of the kernel $Q(\vec{x}, \vec{x}^{\prime})$ for several data points. 
 For the evaluation, we use the classical shadow  \cite{huang2020predicting} as proposed in \cite{Huang2021-pq}.

The resulting performance of the machine learning tasks is shown in TABLE~\ref{table:performance-inner-product}. 
We show the value of mean squared error (MSE) in the case of the regression, and the classification accuracy in the case of the classification, computing from the regression/classification results of the test data. In the same table, we also show the kernel approximation quality $d_Q$. We perform three trials of experiments and compute the mean and the standard deviation of the mean. Note that the number of features in DQF is fixed to the dimension of the Hilbert space where $\rhoq{\vec{x}}$ exists, i.e., 16 in the experiment with the wine quality dataset and 64 in the experiment with the magic04 dataset. In the case of Nystrom, however, we can freely choose the number of features; we choose 16 or 100 in the case of the wine quality dataset and 64 or 100 in the case of the magic04 dataset. 

We see that DQF successfully approximates the quantum kernel in both tasks; $d_Q$ is almost $0.01$ in both experiments. 
Such a good kernel approximation quality results in the performance of the regression/classification tasks. 
DQF outperforms the Nystrom method in MSE/Accuracy. Even if we change the number of features, the performance of the Nystrom method is worse than that of DQF, though the kernel approximation quality improves compared to the one with fewer features. The above results show the validity of using DQF as an alternative to the exact inner product kernel.

\subsection*{Experiment using RQF}

\begin{table*}[ht]
\caption{MSE/Accuracy and the Kernel approximation error $d_Q$ in the regression problem with wine quality dataset in case of the inner product kernel.} 
\label{table:performance-shift}
\begin{tabular}{p{0.15\textwidth} p{0.25\textwidth} p{0.20\textwidth} p{0.20\textwidth} p{0.20\textwidth}}
\hline\hline
Dataset & Algorithm & \#Features  & $d_Q$ & MSE/Accuracy \\
\hline
\multirow{4}{*}{wine quality} &{RQF}   & $100$  &  $0.021\pm 0.002$  & $0.497 \pm 0.002$       \\
&{RQF}   & 200    &  $0.014 \pm 0.002$     &$0.511\pm 0.011$   \\
&{Nystrom} &   100    & $0.007\pm 0.000$    & $0.473 \pm 0.006$        \\
&{Nystrom}  &  200     &   $0.007\pm 0.000$     &      $0.502\pm 0.015$ \\
\hline
\multirow{4}{*}{magic04} &RQF   & 100    &  $0.020\pm 0.000$     &$85.1 \pm  0.0\%$     \\
&{RQF}   & 200    &  $0.015 \pm 0.000$     &$85.5 \pm 0.0\%$ \\
&{Nystrom} &   100     & $0.007\pm 0.000$  & $85.2 \pm 0.0\%$       \\
&{Nystrom}  &  200     & $0.007 \pm 0.000$  & $85.5 \pm 0.0\%$         \\
\hline       
\end{tabular}
\end{table*}

In this experiment, we use the following distance kernel,
\begin{equation}
\label{eq:gauss-experiment}
		Q(\vec{x}, \vec{x}^{\prime}) =  
		\sum_{\vec{q} \in \Lambda} \exp\left(-0.1 ||\rhoq{\vec{x}} - \rhoq{\vec{x}^{\prime}} ||_{HS} \right), 
\end{equation}
where we set $\Lambda = \{
\{j \}
\}_{j=1}^{11}$ 
in the experiment with the wine quality dataset and $\Lambda = \{\{2j -1, 2j\}\}_{j=1}^{5}$ in the experiment with the magic04 dataset. The corresponding feature vector for the quantum kernel \eqref{eq:gauss-experiment} is RQF introduced in Section~\ref{section:feature-representation-shift}. The number of measurements for building each element of the feature vector is set to be $\ns=500$. 

Same as the experiment with DQF, 
we also perform an experiment with the Nystrom method \cite{williams2000using}. For the evaluation of $Q(\vec{x}, \vec{x}^{\prime})$, which is required in the Nystrom method as its subroutine, we estimate each element of $\rhoq{\vec{x}}$ with $\ns=500$ measurements and build $Q(\vec{x}, \vec{x}^{\prime})$ according to \eqref{eq:gauss-experiment}. 

The resulting performance of the machine learning tasks is shown in TABLE~\ref{table:performance-shift}. Same as the previous experiment, we perform three trials of experiments and compute the mean and the standard deviation of the mean. Unlike the previous experiment using DQF, we see that not only RQF but the Nystrom method successfully approximates the quantum kernel, and RQF does not outperform the Nystrom method. 

Still, the results do not directly mean that we should use the Nystrom method instead of RQF. As we discuss at the end of Section~\ref{section:summary-discussion-features}, we can flexibly tune RQF. One of such tunings is embedding parameters in RDM and optimizing parameters. Another possible tuning is changing the way of sampling elements of RQF. In the context of the random Fourier feature, there are many proposals to optimize the way of sampling elements \cite{Rahimi2008-co,Le2013-hx,Wilson2013-ws,Dai2014-vm,Yang2014-po,Feng2015-kd,Yang2015-pt,Yu2015-ko,Choromanski2016-gc,Sinha2016-mm,Oliva2016-my,Yu2016-cl,Lyu2017-ue,Choromanski2017-bd,Li2017-ds,Shen2017-fi,Dao2017-km,Chang2017-ba,Bullins2017-gq,Munkhoeva2018-rk,Rudi2018-vq,Shahrampour2018-ms,Li2019-sj,Agrawal2019-rd,Li2019-mn,Shen2019-ae,Zhang2019-hp,Liu2020-mo,Erdelyi2020-kf}. Studying the effects of tuning RQF is a good direction for future work.


\section{Discussion}

The quantum kernel method is actively examined as an algorithm that may have a quantum advantage, but the generalization capability issue and the scalability issue severely limit its range of application. 
Previous literature \cite{Huang2021-pq,Kubler2021-pm} proposed the quantum kernel with generalization capability (QKGC) using the reduced density matrix (RDM), but the ways of solving the scalability issue remained to be solved. 

In this paper, we first examined classes of QKGC other than the ones using RDM. We examined a quantum kernel with reduced observable (RO), but our theoretical analysis showed that the quantum kernel with RO is almost equivalent to the quantum kernel with RDM. We also examined naively constructible QKGCs, but they lack quantum advantage. From this observation, we focused on solving the scalability issue of QKGC with RDM. 
We then showed that the quantum kernel using QKGC with RDM is almost equivalent to the inner product of two feature vectors, DQF and RQF; as a result, the machine learning problem boils down to a linear problem that can be efficiently solved. 
We also showed that in a wide range of problem settings, the linear model with those features overcomes the scalability issue as in Fig.~\ref{fig:contour}. 

We consolidated the validity of using DQF by numerical experiments compared with a kernel approximation technique called the Nystrom method. DQF successfully outperforms the Nystrom method in regression and classification tasks using a dataset with $O(1,000) \sim O(10,000)$. 
On the other hand, RQF did not outperform the Nystrom method with the same dataset. Still, since there is a large room for tuning RQF, we need further investigation regarding how to utilize RQF.

In our work, we showed that currently conceivable QKGCs can be written using feature vectors and are free from the scalability issue, which leads us to have an opposite emphasis. Namely, we may say that, {\it if we propose a new QKGC, it is preferable to be written using a feature vector}. This constraint will be a good guideline for building a new QKGC.

We also believe that the feature representations can be utilized to recapture the quantum kernel method beyond the tool for solving the scalability issue. For example, as discussed at the end of Section~\ref{section:summary-discussion-features}, we can relate the data reuploading model with the quantum kernel method by embedding the parameters in DQF and RQF and optimizing them. As another example, we may derive the condition for obtaining a quantum advantage as a constraint for the feature vectors by reexamining the discussion in \cite{Huang2021-pq} in light of DQF and RQF. We believe those understandings using the feature representations will pave the way for realizing quantum machine learning algorithms with an advantage over their classical counterpart.

\begin{acknowledgements}
This work was supported by Grant-in-Aid for JSPS Research Fellow 22J01501 and 
MEXT Quantum Leap Flagship Program Grants No. JPMXS0118067285 and No. JPMXS0120319794.
\end{acknowledgements}
\bibliographystyle{unsrt}
\bibliography{main}


\newpage
\onecolumngrid

\appendix
\section{Review of the random Fourier feature} 
\label{section:random-fourier-feature}
Machine learning with random Fourier features \cite{Rahimi2007-om} is a widely used technique to solve the scalability issue of the kernel method. Random Fourier features technique is available when the kernel function $Q(\vec{x}, \vec{x}^{\prime})$ with $\vec{x}, \vec{x}^{\prime} \in \mathcal{X} \subset \r^d$ is written in the form of $Q(\vec{x}, \vec{x}^{\prime}) = Q(\vec{x} - \vec{x}^{\prime})$. The theoretical foundation of random Fourier features is given by Bochner's theorem \cite{Bochner2020-et}, which states that a positive definite distance kernel on $\r^d$ can be written in the form of the Fourier transform of a non-negative measure. Particularly, if the kernel is properly scaled as $Q(\vec{0}) = 1$, 
the kernel is written as 
\begin{equation}
\label{eq:botchner}
	Q(\vec{x} - \vec{x}^{\prime}) = \int_{\r^d} d\vec{\omega} p(\vec{\omega}) e^{i\vec{\omega}^T(\vec{x} - \vec{x}^{\prime})},
\end{equation}
with $p(\vec{\omega})$ as a probability distribution on $\r^d$ defined by the Fourier transform of $Q(\vec{\Delta})$ as 
\begin{equation}
	p(\vec{\omega}) = \frac{1}{(2\pi)^d} \int_{\r^d} d\vec{\Delta} Q(\vec{\Delta}).
\end{equation}
Note that the kernel is a real-valued function, and therefore, the imaginary part of \eqref{eq:botchner} disappears; namely, 
$Q(\vec{x}-\vec{x}^{\prime}) = \int_{\r^d}p(\vec{\omega})\cos(\vec{\omega}^T(\vec{x}-\vec{x}^{\prime}))$. 
Thus by properly choosing a probability distribution $p(\vec{\omega})$, 
\begin{equation}
\begin{split}	
\label{eq:rff-sample}
	Q(\vec{x} - \vec{x}^{\prime}) &= \int_{\r^d}d\vec{\omega} p(\vec{\omega})\cos(\vec{\omega}^T(\vec{x}-\vec{x}^{\prime})) \\
	&= \int_{\r^d} d\vec{\omega} p(\vec{\omega}) \left(
	\cos(\vec{\omega}^T \vec{x}) 	\cos(\vec{\omega}^T \vec{x}^{\prime}) +
	\sin(\vec{\omega}^T \vec{x}) 	\sin(\vec{\omega}^T \vec{x}^{\prime})
	\right)
\end{split}
\end{equation}
holds. 
Sampling $\{\vec{\omega}_1, \vec{\omega}_2,\cdots, \vec{\omega}_{D/2} \}$ from $p(\vec{\omega})$, we can build the random Fourier features as follows:
\begin{equation}
	\vec{z}(\vec{x}) = \sqrt{\frac{2}{D}}
	\left(
	\begin{array}{c}
		\vec{z}_{\vec{\omega}}^c(\vec{x}) \\
		\vec{z}_{\vec{\omega}}^s(\vec{x})
	\end{array}
	\right),~{\rm where}~
	\vec{z}_{\vec{\omega}}^c(\vec{x}) =
	\left(
	\begin{array}{c}
	\cos(\vec{\omega}_1^T \vec{x})\\
	\cos(\vec{\omega}_2^T \vec{x})\\
	\vdots \\
	\cos(\vec{\omega}_{D/2}^T \vec{x})
	\end{array}
	\right),~{\rm and}~
		\vec{z}_{\vec{\omega}}^s(\vec{x}) =
	\left(
	\begin{array}{c}
	\sin(\vec{\omega}_1^T \vec{x})\\
	\sin(\vec{\omega}_2^T \vec{x})\\
	\vdots \\
	\sin(\vec{\omega}_{D/2}^T \vec{x})
	\end{array}
	\right).
\end{equation}
Then, 
\begin{equation}
\label{eq:monte-carlo}
	Q(\vec{x} - \vec{x}^{\prime}) \simeq \vec{z}(\vec{x})^T \vec{z}(\vec{x}^{\prime}).
\end{equation}

In \cite{Sutherland2015-gq}, it is shown that the probability of the error of the kernel approximation is bounded as 
\begin{equation}
\label{eq:rff-bound}
\begin{split}	
    {\rm Prob}\left[{\rm sup}_{\vec{x},\vec{x}^\prime \in \mathcal{X}}|\vec{z}(\vec{x})^T \vec{z}(\vec{x}^{\prime}) - Q(\vec{x} - \vec{x}^\prime) | > \epsilon\right] 
    \leq 66 \left(\frac{\sigma_p {\rm diam}(\mathcal{X})}{\epsilon}\right)^2 \exp\left(- \frac{D\epsilon^2}{8(d+2)}\right), 
\end{split}	
\end{equation}
where $\sigma_p^2 = \expect{\vec{\omega}\sim p(\vec{\omega})}{\vec{\omega}^T\vec{\omega}}$ and ${\rm diam}(\mathcal{X})$ is the diameter of the space $\mathcal{X}$. Namely, the error scales as $\epsilon \sim d/\sqrt{D}$. 
See also literature \cite{Sriperumbudur2015-lr,Avron2017-uz,Honorio2017-ma,Choromanski2018-ct,Zhang2019-hp} for the updated bounds.

\section{A merit of the data reuploading model}
\label{section:merit-data-reuploading}
In this Section, we show the merit of using the data reuploading model according to the literature \cite{schuld2021effect}. 
For simplicity, let us consider the case where the dimension of the data is one and 
	\begin{equation}
	\label{eq:naive-encoding}
		\ket{\psi(\vec{\theta}, x)} = U(\vec{\theta}) 
		\exp(i V x), 
	\end{equation}
	with $V$ as a Hermitian operator with two eigenvalues $\pm \lambda$. Then the model function constructed by $f_{\vec{\theta}}(x) = \bra{\psi(\vec{\theta}, x)}O\ket{\psi(\vec{\theta}, x)}$ with \eqref{eq:naive-encoding}, is limited to the form of
	\begin{equation}
		f_{\vec{\theta}}(x) = c_0(\vec{\theta}) + c_1(\vec{\theta}) e^{i\lambda x} + {\rm c.c.},
	\end{equation}
	where $c_0(\vec{\theta})$ and $c_1(\vec{\theta})$ are the function only dependent on $\vec{\theta}$ \cite{schuld2021effect}. 
	
	To the contrary, if we use the data reuploading model function constructed by setting $U_{\ell}(\vec{x}) = \exp(i V x)\ (\forall \ell)$ in \eqref{eq:data-reuploading}, we obtain
	\begin{equation}
		f_{\vec{\theta}}(x) = \sum_{\omega \in \Omega} c_{\omega} e^{i\omega x},
	\end{equation}
	where the set $\Omega$ is defined by 
	\begin{equation}		
	\Omega = \{
	\omega | (p-q)\lambda~{\rm with}~p, q \in {\bf N}~{\rm and}~1\leq p, q \leq L\}.
	\end{equation}
	Namely, the expressibility of the QNN model is limited because the number of Fourier components is small; the data reuploading model gives the solution to the problem \cite{schuld2021effect}. 

\section{Analysis regarding QKGC with RO}
\label{section:analysis-ro}
Let us define  
\begin{equation}
O_{j\ell}^{ab}(U(\vec{x})) \equiv \bra{a}
{\rm Tr}_{\neq \vec{q}}
	\left(
	U(\vec{x})P_j U(\vec{x})^{\dagger} \rho U(\vec{x})P_{\ell} U(\vec{x})^{\dagger} \right) \ket{b}.
\end{equation} 
We show that $O_{j\ell}^{ab}(U(\vec{x}))$ vanishes if $j \neq \ell$ for almost all of $U(\vec{x})$. For the purpose, we investigate the behavior of $O_{j\ell}^{ab}(U(\vec{x}))$ when $U(\vec{x})$ is distributed according to the Haar measure. 
More specifically, we compute the first two moments of $O_{j\ell}^{ab}(U)$: 
\begin{align}
\label{eq:first-moment}
	\langle O_{j\ell}^{ab} \rangle_{\rm Haar} &\equiv \int_{\rm Haar} dU O_{j\ell}^{ab}(U), \\
\label{eq:second-moment}
	\langle O_{j\ell}^{ab} O_{j\ell}^{ab \ast} \rangle_{\rm Haar} &\equiv \int_{\rm Haar} dU O_{j\ell}^{ab}(U) O_{j\ell}(U)^{ab \ast}. 
\end{align}

For the first moment \eqref{eq:first-moment}, we can use the element-wise integration formula \cite{puchala2011symbolic}
\begin{equation}
\int_{\rm Haar} U_{ab} U^{\ast}_{cd}U_{ef} U^{\ast}_{gh} = \frac{1}{2^{2n}-1}\left(\delta_{ac}\delta_{bd}\delta_{eg}\delta_{fh}+\delta_{ag}\delta_{bh}\delta_{ce}\delta_{df} \right)\nonumber 
-\frac{1}{2^n (2^{2n}-1)}(\delta_{ac}\delta_{bh}\delta_{eg}\delta_{fd} + \delta_{ah}\delta_{bd}\delta_{ec}\delta_{fg}),
\end{equation}
where $n$ is the number of qubits in $U$. A direct calculation shows 
\begin{equation}
\label{eq:result-first-moment}
	\langle O_{j\ell}^{ab} \rangle_{\rm Haar} =  \frac{2^{2n-m}}{2^{2n} - 1}\delta_{j\ell} \delta_{ab} + O\left(\frac{1}{2^n}\right), 
\end{equation}
where $m$ is the number of qubits in $\vec{q}$.

For the second moment \eqref{eq:second-moment}, we use the Weingarten calculus \cite{collins2006integration}, 
\begin{equation}
	U_{i_1 j_1}	U_{i_2 j_2}	U_{i_3 j_3} U_{i_4 j_4} 	
	U_{i_1^{\prime} j_1^{\prime}}^{\ast}	U_{i_2^{\prime} j_2^{\prime}}^{\ast}
		U_{i_3^{\prime} j_3^{\prime}}^{\ast} U_{i_4^{\prime} j_4^{\prime}}^{\ast}
		= \sum_{\sigma, \tau \in S_4} 
		\delta_{i_1 i_{\sigma(1)}^{\prime}} \delta_{i_2 i_{\sigma(2)}^{\prime}} 
		\delta_{i_3 i_{\sigma(3)}^{\prime}} \delta_{i_4 i_{\sigma(4)}^{\prime}} 
		\delta_{j_1 j_{\tau(1)}^{\prime}} \delta_{j_2 j_{\tau(2)}^{\prime}} 
		\delta_{j_3 j_{\tau(3)}^{\prime}} \delta_{j_4 j_{\tau(4)}^{\prime}} 
		{\rm Wg}(\sigma \tau^{-1}),
\end{equation}
where $S_4$ is the set of the symmetric group with degree four and ${\rm Wg}(\cdot)$ is the Weingarten function \cite{collins2006integration} that takes an element of the symmetric group as its input. A direct but complicated calculation shows
\begin{equation}
\label{eq:result-second-moment}
	\langle O_{j\ell}^{ab} O_{j\ell}^{ab \ast} \rangle_{\rm Haar} 
	= 2^{4n - 2m} {\rm Wg}([1,1,1,1]) \delta_{j\ell} \delta_{ab} + O\left(\frac{1}{2^n}\right), 
\end{equation}
where 
\begin{equation}
	{\rm Wg}([1,1,1,1]) = 
	\frac{2^{4n} - 2^{2n} + 6}{2^{8n} - 14 \times 2^{6n} + 49 \times 2^{4n} - 36 \times 2^{2n}}
\end{equation}
in our notation.

The above computation of the first moment \eqref{eq:result-first-moment} and the second moment  \eqref{eq:result-second-moment} shows that if $j \neq \ell$, the first and the second moment of $O_{j\ell}^{ab}(U)$ is exponentially suppressed by the factor $O(1/2^n)$, meaning that for almost all $U$, $O_{j\ell}^{ab} = O(1/2^n)$. Note that as far as $m$ is taken to be small, the first and the second moment of $O_{j\ell}^{ab}(U)$ with $j=\ell$ is not exponentially suppressed, which is the effect of reducing the dimension of the observable.

\section{Other possible QKGC}
\label{section:QKGC-others-example}

One naive approach to realizing QKGC without using RDM is embedding each data with a tiny coefficient. More concretely, given the dimension of data as $d$, and $x_k$ as a $k$-th component of $\vec{x}$, let us define
\begin{equation}
	\ket{\psi(\vec{x})}  = \prod_{k=1}^d V_k \exp\left(i \epsilon A_k x_k \right) \ket{0}^{\otimes n},
\end{equation}
where $\{V_k\}_{k=1}^d$ are unitary operators, $\{A_k\}_{k=1}^D$ are Hermitian operators, and $\epsilon$ is a tiny coefficient. 
If we define $Q(\vec{x}, \vec{x}^{\prime}) = |\langle\psi(\vec{x})|\psi(\vec{x}^{\prime})\rangle|^2$, we can readily show that 
\begin{equation}
\label{eq:small-coefficient-kernel}
	Q(\vec{x}, \vec{x}^{\prime}) = 1 - \epsilon \sum_{k=1}^d \beta_k (x_k + x_k^{\prime}) + O(\epsilon^2). 
\end{equation}
with $\{\beta_k\}_{k=1}^d$ as real $O(1)$ coefficients. As long as $\epsilon$ is tiny, the quantum kernel defined by \eqref{eq:small-coefficient-kernel} has a value large value even if $\vec{x}$ and $\vec{x}^{\prime}$ are different, and therefore, the QKM with \eqref{eq:small-coefficient-kernel} may have the generalization capability. 

Another approach to realizing QKGC is embedding each data so that the number of Fourier components of the quantum kernel is small. In general, $Q(\vec{x}, \vec{x}^{\prime})$ can be written as 
\begin{equation}
	Q(\vec{x}, \vec{x}^{\prime}) = \sum_{\vec{\lambda}, \vec{\lambda}^{\prime} \in \Lambda} c_{\vec{\lambda}, \vec{\lambda}^{\prime}} e^{i\vec{\lambda}^T \vec{x} + i \vec{\lambda}^{\prime T} \vec{x}} + const, 
\end{equation}
where $c_{\vec{\lambda}, \vec{\lambda}^{\prime}} = c_{\vec{\lambda}, \vec{\lambda}^{\prime}}^{\ast}$
$\Lambda$ as a set of vectors with the dimension of the data and $c_{\vec{\lambda}, \vec{\lambda}^{\prime}}$ are complex coefficients. If the dimension of data and the number of vectors in $\Lambda$ are small, $Q(\vec{x}, \vec{x}^{\prime})$ tends to be large and may have the generalization capability even if the dimension of the Hilbert space is exponentially large. 
For example, if the dimension of data is one and $\Lambda$ = $\{1, -1\}$, the kernel is given by
\begin{equation}
	Q(x, x^{\prime}) = 2 c_{1,1} \sin(x + x^{\prime} + \phi_{1,1}) + 2 c_{1,-1} \sin(x - x^{\prime} + \phi_{1,-1}) + const. , 
\end{equation} 
where $\phi_{1,1}$ and $\phi_{1,-1}$ are real numbers. Obviously, $Q(x, x^{\prime})$ can take a value close to one even if $x$ and $x^{\prime}$ is different.
We can naturally construct quantum kernels with a small number of Fourier components as long as the dimension of data is small and we do not encode data repeatedly. 
  
 However, in utilizing the above-introduced QKGCs, the role of the quantum device is quite limited. In the case of the kernel with tiny coefficients, once we determine the values of $\{\beta_k\}_{k=1}^d$ by a quantum device, we can execute all the other processes by a classical computer. In the case of the kernel where the number of Fourier components is small, we can classically pre-determine the set of frequencies $\Lambda$, and determine the values of 
  $\{c_{\vec{\lambda}, \vec{\lambda}^{\prime}}\}_{\vec{\lambda}, \vec{\lambda}^{\prime}\in \Lambda}$ according to the procedure proposed in \cite{schreiber2022classical}. Then, once they are determined, no quantum computation is necessary. From these observations, we contend that the above QKGCs do not have a quantum advantage, and we do not discuss them further in the main text.

\section{Proof of theorems for the feature representation of the projected quantum kernel}
\label{section:proof-theorem-projected}
\subsection{Proof of Theorem \ref{th:quantum-theorem}}
\begin{proof}
Let us expand each density matrix as 
	\begin{equation}	
	\label{eq:expansion}
		\rhoq{\vec{x}} = \sum_{j=1}^{\Nq^2} c_j(\vec{x}) P_j^{(\vec{q})},
	\end{equation}
with $\{c_j(\vec{x})\}_{j=1}^{\Nq^2}$ as coefficients dependent on $\vec{x}$.
Then, the left-hand side of \eqref{eq:quantum-theorem} becomes, 
\begin{align}
\label{eq:theorem-left-hand-side}
    Q(\vec{x}, \vec{x}^{\prime})	= 
    \sum_{j=1}^{\Nq^2} c_j(\vec{x}) c_{j^{\prime}}(\vec{x}^{\prime})
    {\rm Tr}(\pqj \pqjprime) = \Nq   \sum_{j=1}^{\Nq^2} c_j (\vec{x}) c_j(\vec{x}^{\prime}).
\end{align}
The right-hand side becomes 
\begin{align}	
\label{eq:theorem-right-hand-side}
& \frac{1}{\Nq}\sum_{j=1}^{\Nq^2} {\rm Tr}\left(\pqji \rho({\bf x}) \right) {\rm Tr}\left(\pqji \rho({\bf x}^{\prime}) \right) \\
&=  \frac{1}{\Nq}\sum_{j=1}^{\Nq^2} {\rm Tr}\left( \pqj \rho^{(\vec{q})}(\vec{x})\right) 
{\rm Tr}\left( \pqj \rho^{(\vec{q})}(\vec{x}^{\prime})\right) 
\\
&= \frac{1}{\Nq}\sum_{j,k,\ell=1}^{\Nq^2} c_k(\vec{x}) c_{\ell}(\vec{x}^{\prime})
{\rm Tr}\left(P_j^{(\vec{q})} P_k^{(\vec{q})} \right) {\rm Tr}\left(P_j^{(\vec{q})} P_{\ell}^{(\vec{q})} \right) \\
&= \Nq \sum_{j=1}^{\Nq^2} c_j(\vec{x}) c_{j}(\vec{x}^{\prime}), 
\end{align}
which is equal to the left hand side \eqref{eq:theorem-left-hand-side}.
\end{proof}

\subsection{Proof of the bound for $\ns$}
\label{section:shot-noise}
 Let $\tildepq(\vec{x})$ be an unbiased estimator of ${\rm Tr}\left(\pqji \Phi\left(\rho(\vec{x})\right)\right)$, whose value is determined by $\ns$ measurements (each element of feature $z_j(\vec{x})$ is directly constructed as $z_j(\vec{x}) = \tildepq(\vec{x})/\sqrt{N_{\vec{q}}}$). 
Then we can readily show that if at least
\begin{equation}	
\label{eq:prob-necessary}
|\tildepq(\vec{x}) - {\rm Tr}(\pqji \Phi\left(\rho(\vec{x}))\right)| \leq \frac{\epsilon}{3\Nq} ,
\end{equation}
for all $j$ with  $\epsilon < 1$, 
\begin{equation}
	\left|
	\sum_{j=1}^{\Nq^2} \vec{z}(\vec{x})^T \vec{z}(\vec{x}^{\prime}) 
	- {\rm Tr}\left(\rhoq{\vec{x}}\rhoq{\vec{x}^{\prime}}\right)  
	\right| < \epsilon, 
\end{equation}
is satisfied.
From Hoeffding's inequality, the probability that \eqref{eq:prob-necessary} is not satisfied is bounded as
\begin{equation}
	{\rm Prob}\left(\left|\tildepq (\vec{x}) - {\rm Tr}\left(\pqji\Phi\left(\rho(\vec{x})\right)\right)\right| \geq \frac{\epsilon}{3\Nq} \right)
	\leq 2\exp\left(-
	\frac{\ns \epsilon^2}{18\Nq^2}
	\right).
\end{equation}
Thus, the probability $\delta$ that 
at least one of the inequality \eqref{eq:prob-necessary} is not satisfied in the training data is bounded as
\begin{equation}
	\delta \leq 2M  \exp\left(-
	\frac{\ns \epsilon^2}{18\Nq^2}
	\right).
\end{equation} 
Conversely, if
\begin{equation}
\ns \geq \frac{18\Nq^2}{\epsilon^2} \log\left(\frac{2M}{\delta}\right),
\end{equation}
is satisfied, 
\begin{equation}
	{\rm Prob}\left[{\rm sup}_{\vec{x},\vec{x}^\prime \in \mathcal{X}_{\rm data}}
	\left|
	\vec{z}(\vec{x})^T \vec{z}(\vec{x}^{\prime}) -
	{\rm Tr}(\rhoq{\vec{x}}\rhoq{\vec{x}^{\prime}}) \right| \geq \epsilon\right] 
	\leq \delta
\end{equation}
holds for all data.

\section{Proof of theorems for the feature representation of the distance kernel with RDM}
\label{section:proof-theorem-distance}
\subsection{Proof of theorem~\ref{th:hs-equality}}
With the expansion \eqref{eq:expansion}, 
\begin{equation}
\begin{split}	
	\dhs &= \frac{1}{4}{\rm Tr}\left((\rhoq{\vec{x}} - \rhoq{\vec{x}^{\prime}})^2\right) \\
			&= \frac{\Nq}{4}
			\left|\left|
			\vec{c}(\vec{x}) - \vec{c}(\vec{x}^{\prime})
			\right|\right|_2^2, 
\end{split}
\end{equation}
where $\vec{c}(\vec{x}) = \left\{{\rm Tr}\left(\pqjki \Phi\left(\rho(\vec{x})\right)\right)/\Nq \right\}_{j=1}^{\Nq^2}$. 
Therefore, $f(\dhs)$ is a shift-invariant function in the sense that
\begin{equation}	
f\left(\dhs\right) = F\left(\vec{c}(\vec{x}) - \vec{c}(\vec{x}^{\prime})\right), 
\end{equation}
where $F\in {\r^{\Nq^2} \rightarrow \r }$ is defined by 
$F({\bf b}) = f(\Nq||{\bf b}||_2^2/4)$.

Then, from \eqref{eq:rff-sample}, 
\begin{equation}
\label{eq:proof-hs-equality}
\begin{split}	
	F(\vec{c}\left(\vec{x}) - \vec{c}(\vec{x}^{\prime})\right) 
	&= \int_{\r^{\Nq^2}} d\vec{\omega} p(\vec{\omega}) 
	e^{i\vec{\omega}^T \left(\vec{c}(\vec{x}) - \vec{c}(\vec{x}^{\prime})\right)} \\
	&= \int_{\r^{\Nq^2}} d\vec{\omega} p(\vec{\omega}) 
	\left( \cos(\vec{\omega}^T \vec{c}(\vec{x}))\cos(\vec{\omega}^T \vec{c}(\vec{x}^{\prime})) + 
		\sin(\vec{\omega}^T \vec{c}(\vec{x}))\sin(\vec{\omega}^T \vec{c}(\vec{x}^{\prime}))
	\right), 
\end{split}
\end{equation}
where 
\begin{equation}
	p(\vec{\omega}) = \frac{1}{(2\pi)^{\Nq^2}} \int_{\r^{\Nq^2}} F(\vec{\Delta}),
\end{equation}
Eq.~\eqref{eq:proof-hs-equality} is equivalent to \eqref{eq:hs-equality} in Theorem~\ref{th:hs-equality}.

\subsection{Proof of Theorem~\ref{th:hs-error}}
\label{section:proof-distance-error}
We write 
\begin{equation}
	Q(\vec{x}, \vec{x}^{\prime}) = f\left(\dhs \right).
\end{equation}
From the triangle inequality, it holds
\begin{equation}
\label{eq:triangle}
	{\rm sup}_{\vec{x},\vec{x}^\prime \in \dataspace_{\rm data}}|\vec{z}(\vec{x})^T \vec{z}(\vec{x}^{\prime}) - Q(\vec{x}, \vec{x}^\prime) | 
	\leq {\rm sup}_{\vec{x},\vec{x}^\prime \in \dataspace_{\rm data}}| \tilde{Q}(\vec{x}, \vec{x}^{\prime}) - {Q}(\vec{x}, \vec{x}^\prime) | 
	+ {\rm sup}_{\vec{x},\vec{x}^\prime \in \dataspace_{\rm data}}|\vec{z}(\vec{x})^T \vec{z}(\vec{x}^{\prime}) - \tilde{Q}(\vec{x}, \vec{x}^\prime) |, 
\end{equation}
where 
\begin{equation}	
\tilde{Q}(\vec{x}, \vec{x}^{\prime}) = F(\tilde{\vec{c}}(\vec{x}) - \tilde{\vec{c}}(\vec{x}^{\prime}) ),
\end{equation}
with $\tilde{\vec{c}}(\vec{x})$ as an unbiased estimator of $\vec{c}(\vec{x})$, where each element is determined by $\ns$ measurements. 
We see that the first term on the right-hand side of \eqref{eq:triangle} corresponds to the sampling error due to the limited number of samples $D$, and the second term corresponds to the error due to the shot noise.

For the first term, 
\begin{equation}	
\begin{split}	
|\tilde{Q}(\vec{x}, \vec{x}^{\prime}) - Q(\vec{x}, \vec{x}^{\prime})| 
&= \left|
	f\left(\frac{\Nq || \tilde{\vec{c}}(\vec{x}) - \tilde{\vec{c}}(\vec{x}^{\prime})||_2^2}{4}\right)
	- f\left(\frac{\Nq || {\vec{c}}(\vec{x}) - {\vec{c}}(\vec{x}^{\prime})||_2^2}{4}\right)
	\right| \\
&\leq  \frac{N_{\vec{q}}L_f}{4} \left| || \tilde{\vec{c}}(\vec{x}) - \tilde{\vec{c}}(\vec{x}^{\prime})||_2^2
- || {\vec{c}}(\vec{x}) - {\vec{c}}(\vec{x}^{\prime})||_2^2 
\right| \\
&\leq \frac{N_{\vec{q}}L_f}{4}
\left|
|| \tilde{\vec{c}}(\vec{x}) - \vec{c}(\vec{x}) 
- \left(\tilde{\vec{c}}(\vec{x}^{\prime}) - \vec{c}(\vec{x}^{\prime})\right) 
+ \left(\vec{c}(\vec{x}) -\vec{c}(\vec{x}^{\prime}) \right)    ||^2_2  
- || {\vec{c}}(\vec{x}) - {\vec{c}}(\vec{x}^{\prime})||_2^2  
 \right|
\\
&\leq \frac{N_{\vec{q}}L_f}{4} \left( 
|| \tilde{\vec{c}}(\vec{x}) - {\vec{c}}(\vec{x})||_2^2 +
|| \tilde{\vec{c}}(\vec{x}^{\prime}) - {\vec{c}}(\vec{x}^{\prime})||_2^2 
\right),
\end{split}
\end{equation}
where in the second line, we use that the Lifshitz constant of $f$ with respect to the $L_1$ norm is $L_f$. Thus, if 
\begin{equation}
\label{eq:diff-bound}
	|| \tilde{\vec{c}}(\vec{x}) - {\vec{c}}(\vec{x})||_2^2 < \frac{\epsilon}{\Nq L_f}, 
\end{equation}
for all $\vec{x} \in \mathcal{X}$, 
\begin{equation}
{\rm sup}_{\vec{x},\vec{x}^\prime \in \mathcal{X}_{\rm data}}|\tilde{Q}(\vec{x}, \vec{x}^{\prime}) - Q(\vec{x}, \vec{x}^{\prime})| < \frac{\epsilon}{2}.
\end{equation}
The probability that \eqref{eq:diff-bound} does not hold is bounded from Hoeffding's inequality as 
\begin{equation}
 {\rm Prob}\left[ || \tilde{\vec{c}}(\vec{x}) - {\vec{c}}(\vec{x})||_2^2 \geq \frac{\epsilon} {\Nq L_f} \right]
\leq 2 \exp\left( -\frac{\ns \epsilon^2 }{8 \Nq^2 L_f^2}\right),
\end{equation}
where we use 
\begin{equation}
	0  \leq || \tilde{\vec{c}}(\vec{x}) - {\vec{c}}(\vec{x})||_2^2 
	\leq 4. 
\end{equation}
Thus, 
\begin{equation}
\label{eq:first-probability}
 {\rm Prob}\left[{\rm sup}_{\vec{x},\vec{x}^\prime \in \dataspace_{\rm data}} || \tilde{Q}(\vec{x}, \vec{x}^{\prime}) - {Q}(\vec{x}, \vec{x}^{\prime})||_2^2 \geq \frac{\epsilon} {2} \right]
\leq 2 M \exp\left( -\frac{\ns \epsilon^2 }{8 \Nq^2 L_f^2}\right).	
\end{equation}

Next, we bound the second term in \eqref{eq:triangle}. Given the space where $\vec{c}(\vec{x})$ is located as $\mathcal{X}_{\vec{c}}$, using \eqref{eq:rff-bound} with $d = \Nq^2$, $\mathcal{X} = \mathcal{X}_{\vec{c}}$, and $\epsilon\rightarrow \epsilon/2$, we obtain
\begin{equation}
\label{eq:second-probability}
	{\rm Prob}\left[{\rm sup}_{\vec{x},\vec{x}^\prime \in \mathcal{X}_{\rm data}}|\vec{z}(\vec{x})^T \vec{z}(\vec{x}^{\prime}) - \tilde{Q}(\vec{x}, \vec{x}^\prime) |  \geq \frac{\epsilon}{2}\right] 
    \leq 264 \left(\frac{\sigma_p {\rm diam}(\mathcal{X}_{\vec{c}})}{\epsilon}\right)^2 \exp\left(- \frac{D\epsilon^2}{32(\Nq^2+2)}\right). 
\end{equation}  
Since $|\vec{c}(\vec{x})| \leq 1$, it holds ${\rm diam}(\mathcal{X}_{\vec{c}}) \leq 2$. 
Therefore, 
\begin{equation}
	{\rm Prob}\left[{\rm sup}_{\vec{x},\vec{x}^\prime \in \mathcal{X}_{\rm data}}|\vec{z}(\vec{x})^T \vec{z}(\vec{x}^{\prime}) - \tilde{Q}(\vec{x}, \vec{x}^\prime) |  \geq \frac{\epsilon}{2}\right] 
    \leq 1056 \left(\frac{\sigma_p}{\epsilon}\right)^2 \exp\left(- \frac{D\epsilon^2}{32(\Nq^2+2)}\right). 
\end{equation}  

From the above, we observe the followings: 
\begin{itemize}
	\item (a) From \eqref{eq:first-probability}, with probability less than or equal to $\delta/2$, 
	\begin{equation}
		{\rm sup}_{\vec{x},\vec{x}^\prime \in \mathcal{X}_{\rm data}}|\tilde{Q}(\vec{x}, \vec{x}^{\prime}) - Q(\vec{x}, \vec{x}^{\prime})| \geq \frac{\epsilon}{2},
	\end{equation}
	if $\ns$ is bounded as
	\begin{equation}
	\label{eq:shot-bound}
		\ns \geq \frac{8\Nq^2 L_f^2}{\epsilon^2} \log\left(\frac{4M}{\delta}\right).
	\end{equation} 
	\item (b) From \eqref{eq:second-probability}, with probability less than or equal to $\delta/2$, 
	\begin{equation}
		{\rm sup}_{\vec{x},\vec{x}^\prime \in \mathcal{X}_{\rm data}}|\vec{z}(\vec{x})^T \vec{z}(\vec{x}^{\prime}) - \tilde{Q}(\vec{x}, \vec{x}^\prime) | \geq \frac{\epsilon}{2}, 
	\end{equation}
	if $D$ is bounded as 
	\begin{equation}
		\label{eq:sample-bound}
		D \geq \frac{32(\Nq^2 + 2)}{\epsilon^2} \log\left(\frac{2112\sigma_p^2}{\epsilon^2 \delta}\right).
	\end{equation}
\end{itemize}
Combining (a) and (b) with Eq.~\eqref{eq:triangle}, we conclude that 
\begin{equation}
	{\rm Prob}\left[ \left|
		{\rm sup}_{\vec{x},\vec{x}^\prime \in \mathcal{X}_{\rm data}}|\vec{z}(\vec{x})^T \vec{z}(\vec{x}^{\prime}) - Q(\vec{x}, \vec{x}^\prime) \right| \geq \epsilon 
	\right] \leq \delta
\end{equation}
if \eqref{eq:shot-bound} and \eqref{eq:sample-bound} are both satisfied.

\section{Detail of the numerical experiment}
\label{section:detail-experiment}
\subsection{Data encoding circuit}
\begin{figure}[ht]
\scalebox{1.0}{
\Qcircuit @C=1.0em @R=0.2em @!R { \\
	 	\nghost{{q}_{1} :  } & \lstick{{q}_{1} :  } & \gate{\mathrm{H}} & \gate{\mathrm{R_Y}\,(\mathrm{x_1})} & \ctrl{1} & \ctrl{2} & \ctrl{3} \barrier[0em]{3} & \qw & \qw & \qw \barrier[0em]{3} & \qw & \qw \barrier[0em]{3} & \qw & \qw & \qw\\
	 	\nghost{{q}_{2} :  } & \lstick{{q}_{2} :  } & \gate{\mathrm{H}} & \gate{\mathrm{R_Y}\,(\mathrm{x_2})} & \targ & \qw & \qw & \qw & \ctrl{1} & \ctrl{2} & \qw & \qw & \qw & \qw & \qw\\
	 	\nghost{{q}_{3} :  } & \lstick{{q}_{3} :  } & \gate{\mathrm{H}} & \gate{\mathrm{R_Y}\,(\mathrm{x_3})} & \qw & \targ & \qw & \qw & \targ & \qw & \qw & \ctrl{1} & \qw & \qw & \qw\\
	 	\nghost{{q}_{4} :  } & \lstick{{q}_{4} :  } & \gate{\mathrm{H}} & \gate{\mathrm{R_Y}\,(\mathrm{x_4})} & \qw & \qw & \targ & \qw & \qw & \targ & \qw & \targ & \qw & \qw & \qw\\
\\ }}
\caption{The data encoding circuit when $n=4$.}
\label{fig:circuit}
\end{figure}
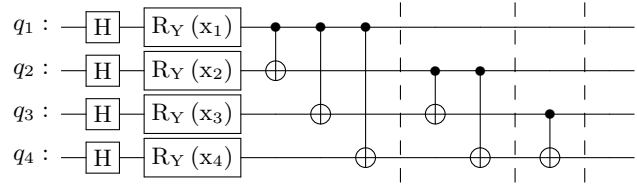

Here let us show the data encoding circuit in the numerical experiment in Section~\ref{section:numerical}. Suppose that the dimension of the data is $n$. The number of qubits is also $n$ (recall that the dimension of the data is equal to the number of qubits in our numerical experiment).
Then, as the data encoding circuit, we use the following unitary operator
\begin{equation}
	U(\vec{x}) = U_{\rm ent} U_Y(\{x_j \}_{j=1}^n) H^{\otimes n}, 
\end{equation}
and build the density operator before the reduction of qubits as $\rho(\vec{x}) = U(\vec{x})|0\rangle^{\otimes n} \langle 0|^{\otimes n} U^{\dagger}(\vec{x})$.
Here, we define 
\begin{equation}
	U_Y(\{x_j\}_{j=1}^n) = R_Y (x_1) \otimes R_Y(x_2) \otimes \cdots 
	\otimes R_Y(x_n). 
\end{equation}
Also, we define
\begin{equation}
	U_{\rm ent} = U_{\rm ent}^{(n)}  \cdots U_{\rm ent}^{(2)} U_{\rm ent}^{(1)},
\end{equation}
with
\begin{equation}
	U_{\rm ent}^{(j)} =   {\rm CNOT}(j, n) \cdots {\rm CNOT}(j, j+2){\rm CNOT}(j, j+1),
\end{equation}
where ${\rm CNOT}(j, k)$ is the CNOT operator with $j$ as the control qubits and $k$ as the target qubits. We show the data encoding circuit when $n=4$ as an example in Fig.~\ref{fig:circuit}.

\subsection{Nystrom method}
The Nystrom method, which we demonstrate in Section~\ref{section:numerical} is another algorithm to construct features that approximates the kernel. The procedure for approximating a kernel $Q(\vec{x}, \vec{x}^{\prime})$ is as follows. 
We first sample $D$ data, where we write them as $\{\hat{\vec{x}}_1, \hat{\vec{x}}_2,\cdots \hat{\vec{x}}_D\}$. 
By using the sampled data, we construct the following matrix $\hat{Q} \in \r^D \times \r^D$ defined by $\hat{Q}=\{Q(\hat{\vec{x}}_k, \hat{\vec{x}}_{\ell})\}$, where the indices $k, \ell$ run from $1$ to $D$. 

Suppose that the non-zero eigenvalues of $\hat{Q}$ as $\{d_s\}_{s=1}^{S}$ and the corresponding eigenvectors as $\{\vec{v}_s\}_{s=1}^{S}$, where $S \leq D$.
Given matrices
\begin{equation}
	A = diag\left(\frac{1}{\sqrt{\lambda_1}}, \frac{1}{\sqrt{\lambda_2}}, \cdots, \frac{1}{\sqrt{\lambda_S}}\right)~{\rm and}~B = (\vec{v}_1, \vec{v}_2, \cdots \vec{v}_S), 
\end{equation}
we build the features as 
\begin{equation}
	\vec{z}(\vec{x}) = A B^T 
	\left(
	\begin{array}{c}
	Q(\vec{x}, \hat{\vec{x}}_1) \\
	Q(\vec{x}, \hat{\vec{x}}_2) \\
	\vdots \\
	Q(\vec{x}, \hat{\vec{x}}_S) \\
	\end{array}
	\right). 
\end{equation}
Note that the kernel $\tilde{Q}(\vec{x}, \vec{x}^{\prime}) = \vec{z}(\vec{x})^T \vec{z}(\vec{x}^{\prime})$ is the approximate of the original kernel $Q(\vec{x}, \vec{x}^{\prime})$.


\end{document}